\documentclass{aa}
\usepackage{natbib,graphicx,ulem}
\usepackage{psfig,graphicx,epsfig}
\usepackage{soul,color}
\usepackage{longtable,lscape}
\usepackage{amsmath}

\DeclareGraphicsExtensions{.pdf,.png,.jpg}

\begin{document}

\title{A metal-rich elongated structure in the core of the group NGC4325}

\author{
T.~F.~Lagan\'a  \inst{1} 
\and L. Lovisari \inst{2}
\and L. Martins \inst{1}
\and G. A. Lanfranchi \inst{1}
\and H. V. Capelato \inst{1,3}
\and G. Schellenberger \inst{2}
}

\institute{
N\'{u}cleo de Astrof\'{i}sica Te\'{o}rica, Universidade Cruzeiro do Sul, Rua Galv\~{a}o Bueno 868, 
Liberdade, CEP: 01506-000, S\~{a}o Paulo, SP, Brazil
\and
Argelander-Institut f\"ur Astronomie, Universit\"at Bonn, Auf
 dem H\"ugel 71, 53121 Bonn, Germany
 \and
 Divis\~{a}o de Astrof\'{i}sica, INPE/MCT, 12227-000, S\~{a}o Jos\'{e} dos Campos/SP, Brazil
}


\authorrunning{Lagan\'a et al.}

\titlerunning{the galaxy group of NGC4325}

\abstract
{}
{Based on XMM-\textit{Newton}, \textit{Chandra}, and optical DR10-SDSS data, we investigate the metal enrichment history of the group 
NGC4325 (z$=0.026$). To complete the analysis we used chemical evolution models and studied the optical spectrum of the central dominant galaxy 
through its stellar population analysis and emission line diagnostics to analyse its role in the  
metal enrichment of the intra-group medium.}
{We used X-ray 2D spectrally resolved maps to resolve structure in temperature and metallicity. We also derived gas and total masses within $r_{2500}$ 
and $r_{500}$ assuming hydrostatic equilibrium and spherical symmetry. 
To perform stellar population analysis we applied the spectral fitting technique with \texttt{STARLIGHT} to the optical spectrum of the central galaxy.
We simulated the chemical evolution of the central galaxy.}
{While the temperature, pseudo-pressure, and pseudo-entropy maps showed no inhomogeneities, the iron spatial distribution shows a filamentary structure 
in the core of this group, which is spatially correlated with the central galaxy, suggesting a connection between the two.
The analysis of the optical spectrum of the central galaxy showed no contribution by any recent AGN activity. 
Using the star formation history as input to chemical evolution models,
we predicted the iron and oxygen mass released by supernovae (SNe) winds in the central galaxy up to the present time. }
{Comparing the predicted amount of mass released by the NGC4325 galaxy 
to the ones derived through X-ray analysis
we conclude that the winds from the central galaxy alone play a minor role in the IGM metal enrichment of this group inside $r_{2500}$. 
The  SNe winds are responsible for no more than
3$\%$ of it and of the iron mass and 21$\%$ of the oxygen mass enclosed within $r_{2500}$.
Our results suggest  that  oxygen has been produced in the early stages of the group formation, 
becoming well mixed and leading to an almost flat profile. Instead, the iron distribution is centrally peaked, indicating
that this element is still being added to the IGM specifically in the core by the SNIa. 
A possible scenario to explain the elongated metal-rich structure in the core of the NGC4325 is a past AGN activity, in which our results suggest
an episode older than $\sim 10^{7}-10^{8}$ yrs and younger than $5 \times 10^{8}$. Through the overall distribution of the galaxies, we found no signs of 
recent merger in the group centre that could explain the metal-rich structure. }


\maketitle

\section{Introduction}

On the basis of the X-ray surface brightness profile of galaxy clusters, one
can distinguish two types of systems: the first  are the clusters 
with a sharp surface brightness peak towards the centre, namely the cool-core (CC)
clusters. The second type are the non-cool-core clusters, which do not show such 
a surface brightness peak. Cool-core clusters are relatively  dynamically relaxed systems
in which the cooling time of the dense X-ray emitting plasma in the central region is
short compared to the Hubble time, which leads to the 
cooling-flow model \citep[e.g.][]{fabian94}. As a result, large amounts of cold gas and star formation
are expected to be found at the centre of these systems.
The first high-resolution X-ray spectra of clusters of galaxies
taken with the reflection grating spectrometers (RGS) of
XMM-\textit{Newton}  indeed showed cooler gas in the
cores of several clusters. However, the amount of cool gas at
lower temperatures was much less than predicted by the 
cooling flow model \citep{peterson01,tamura01,kaastra01,peterson03}.
The absence of a cool phase in cores of galaxy clusters is suggestive of one or more heating mechanisms maintaining the 
hot gas at keV temperatures. 
Among the many candidate heating mechanisms put forth recently, the most successful one for gas heating in cluster cores 
has been feedback from active galactic nuclei (AGNs), for which observational evidence has been growing in recent 
years \citep[see][for a review]{McNamara07,Gitti12}.

Another indication of non-gravitational energy injection into the intracluster medium (ICM) comes from 
the deviation of scaling relations from self-similar models. The most famous and studied of these deviations is 
between cluster X-ray luminosity  ($L_{\rm bol}$) and X-ray temperature ($T_{\rm X}$),
where observational studies show an exponent closer to 2.9 instead of the predicted value of two 
\citep[e.g.][]{mark98,AE99,pratt09}. The comparison between scaling relations in the purely gravitational 
picture of cosmic structure growth and the observed X-ray scaling relations within a scenario including hydrodynamical effects
are  discussed in detailed in \citet{Bohringer12}.
These deviations from self-similar collapse are thought to be best characterised by the injection of energy
into the gas before cluster collapse \citep[preheating;][]{kaiser91,EK91,Bialek01}. In addition to preheating, internal heating by supernovae and
AGNs may also impact the observed scaling relations \citep[e.g.][]{Tozzi01}.


Additional evidence of non-gravitational energy input into the ICM is the presence of metals. The amount of metals
in the ICM is roughly one-third the solar value, regardless of the cluster size. It is remarkable that the enrichment level is
similar in clusters of all masses, despite the significant trend found for the variation in the star formation efficiency 
\citep{LMS03,giodini09,lagana11,andreon12,lagana13}.  If we consider that a few percent of the baryons are formed 
into stars and that stars in cluster galaxies
can only provide a one-sixth solar metallicity \citep{portinari04}, 
the metal enrichment process represents 
a challenge. Contributing to the difficulty of a complete understanding of metal enrichment history, \citet{werner13} 
have recently presented still another scenario in which most of the metal enrichment of the intergalactic medium 
occurred before the cluster formed, during the period of maximal star formation and black hole activity.

It  has been revealed that the iron abundance distribution in a few bright clusters of galaxies are not central-peaked,
as shown in the Perseus cluster \citep{sanders05} and Abell 1060 \citep{hayakawa06}, among others. 
In these clusters, the high abundances are observed in off-centre regions
located a few kiloparsecs from the the cluster centre.
In the Perseus cluster, the high-abundance region coincides with a radio mini-halo, indicating that it
may be caused by metal-enriched bubbles lifting from the  AGN.
The off-centre metal-rich structure does not especially come from clusters. 
The presence of high-abundance structures has also been reported in groups of galaxies, such as in HCG 62 \citep{gu07} or RGH 80 \citep{cui10}.
The former authors affirm that the high abundance found in HCG 62  could have developed during an episode of AGN activity,
while the latter authors argue in favour of a metal enrichment through ram pressure stripping.

As seen, various processes can contribute to transport enriched gas from galaxies to their environment \citep[for a review see][]{schindler08}:
ram-pressure stripping \citep{GG72}, tidal stripping \citep{Toomre72}, AGN outflows, galactic winds \citep{DeYoung78}, 
supernova explosions \citep[e.g.][]{Veilleux05}, etc. 
The efficiency of the process depends on the 
environmental properties and on the galaxy providing metals.
Thus, the measurement of heavy element abundances in
the intragroup  medium (IGM) can provide important clues to the chemical evolution
inside galaxy groups, because 

different processes of metal 
transportation lead to different metal spatial distributions \citep[e.g.][]{schindler08}. Also, 
since groups of galaxies have smaller potential wells,
non-gravitational processes are expected to be more important and thus, the spatial distribution
of metals in groups can be different from clusters of galaxies.
However, difficulties in determining the nature of 
metal enrichment in clusters and groups are enhanced not only by 
the variety of the astrophysical
processes and spatial scales involved but also by the high uncertainties in the
abundance determination of elements other than iron, especially for low-mass systems. 
Although the analysis of clusters of galaxies has received the bulk of observational attention,
some recent papers added in the metal-enrichment analysis of galaxy groups
 \citep[e.g.][]{dePlaa06,rasmussen09,sato09,sato10,sasaki14}.

Since in a previous work \citet{russell07} suggested a past AGN activity but did no metal-enrichment analysis,
the aim of this paper is to investigate a possible connection with an enhancement of the metallicity
as observed for other objects \citep{Churazov01,Simionescu08,Simionescu09}.
Thus, in this paper, we report an X-ray and optical analysis of the NGC4325 galaxy group,
looking at the spatial distribution of the temperature and the metallicity and searching for
imprints of the metal enrichment history and its connection to the central galaxy.
Since we have performed different analysis (2D spectral resolved maps + stellar population analysis + chemical evolution model 
+ emission lines diagnostic + optical),  we can now give an unrivalled overview that can advance our understanding of the metal 
enrichment history of galaxy groups.

The rest of the paper is organised as follows. After the Introduction, we present 
the X-ray data analysis in Sect.~\ref{xrayana}, followed by the results in Sect.\ref{res}.
In Sect.~\ref{cetralgal} we present the stellar population analysis, emission line mechanisms,  and chemical evolution model of the central galaxy,
followed by our discussions in Sect.~\ref{disc}.
We end with the conclusions in Sect.~\ref{conc}.
We adopt a $\Lambda$CDM cosmology with $H_{\rm 0} = \rm 70 ~ km~ s^{-1} ~Mpc^{-1}$,
$\Omega_{\rm M} = 0.3$, $\Omega_{\rm \Lambda} = 0.7$, and h(z) is the ratio of the Hubble constant
at redshift z to its present value, $H_{0}$. 
Confidence intervals correspond to the 90\% confidence level.

\section{X-ray data analysis}
\label{xrayana}
The XMM-\textit{Newton} observations (ID: 0108860101) were made with the Thin 1 filter in
Prime full window on  December 24, 2000.
Data from the European Photon Imaging Camera (EPIC) were prepared for 
examination using the \texttt{emchain} and   \texttt{epchain} processing routines within
the Science Analysis Software (SAS v12.0.1), thus creating calibrated event files for each detector.
The initial data screening was applied using  recommended sets of
events with \texttt{FLAG}=0 
and \texttt{PATTERN} 0-12 and 0-4 for the MOS and pn cameras, respectively. 

We applied a 2$\sigma$ clipping procedure to filter flares using the light curves
in the [10-12] keV energy band, where the particle background dominates the source counts.
There are 18.3 ks clean data for MOS1, 16.8 ks for MOS2, and 11.9 ks for pn. 
With the cleaned event files, we created the redistribution matrix file (RMF) and ancillary file (ARF)
with the SAS tasks \texttt{rmfgen} and \texttt{arfgen} for each camera and for each region that we analysed. 

To take into account for each detector the background contribution,   we obtained a background spectrum
in an outer annulus of the observation, 12.5-14 arcmin, and in the [10-12]keV energy band.
Then, we compared these spectra with the one obtained
from the \citet{RP03} blank sky  in the same region and energy band.
We thus divided the observation background by the blank sky background to obtain a 
normalization parameter for each detector  that will be used in all spectral fits \citep[as already presented in][]{lagana08}.

We also analysed a \textit{Chandra} ACIS-S
observation (ID: 3232). We performed a standard data
reduction using the CIAO 4.6 analysis tools and the
\textit{Chandra} CALDB 4.6. We first created the level-2 event
file with the \texttt{chandra$\_$repro} task and the
\texttt{lc$\_$clean} algorithm to filter out flared time
intervals. We were able to use 96\% (29ks) of the uncleaned
observation for our analysis. For the background
subtraction, we used the blank-sky background files and
rescaled the corresponding spectra according to the count
rate in the [9.5--12] keV energy band of the
observation.

\subsection{Spectral fitting}
\label{specfit}

For the spectral analysis, the cluster's X-ray emission was modelled with the single
temperature plasma model \texttt{APEC} in  \texttt{XSPEC} v12 using the \texttt{AtomDB} v2.0.2,
which includes major updates, specifically for the Fe-L and Ne-K complex lines \citep{foster10}.
These changes are critical for studying the metallicity in low-temperature systems, such as galaxy groups 
where the Fe-L lines dominate the spectra. A detailed analysis of the effects of the \texttt{AtomDB} updates
in the temperature and abundances of low-mass systems is discussed in \citet{lovisari14}. \\

All the spectra were fitted in the [0.5-5] keV band, with the low-energy cut-off 
chosen such as to reduce contamination from the residual Galactic soft emission. Above
5 keV we do not expect any emission from this low-temperature group, and therefore we increase the S/N with this cut.
We also excluded the energy band from 1.3 to 1.9 keV to avoid any influence of the strong Al and Si lines.

For each fit, the source redshift was fixed to $z=0.0257$, and the hydrogen column density, $n_{\rm H}$, was 
frozen to the galactic value of $2.32 \times 10^{20} \rm cm^{-2}$, obtained using the task \texttt{nh} of \texttt{FTOOLS}  \citep[based on][]{lab05}.
The gas temperature kT, metallicity Z, and normalization
$N_{\rm apec}$ (effectively the emission measure) were allowed to vary in the \texttt{APEC} plasma model.
Also, the normalization of pn was free to vary (not linked to the MOS).
The metallicity were based on the solar values given by \citet{aspl09}.

Just for the iron and oxygen profiles presented in Sect.~\ref{res} we extracted spectra in concentric annuli using \texttt{VAPEC} plasma model,  
centred on the peak of diffuse X-ray emission, 
with a minimum width of 30 arcsec and containing at least 4500 cts, which allowed us to estimate the oxygen abundance with a relatively good accuracy.
In this case, we also fixed the He, C and N abundances to the solar value.

\subsection{2D spectral mapping}

To resolve structures in temperature and metallicity,
we divide the data into small regions from which spectra can be
extracted.  The 2D maps
were made in a grid,  
where each pixel; is 512 $\times$ 512 XMM EPIC physical pixels, i.e., each
cell grid is 25.6 arcsec $\times$ 25.6 arcsec.
The grid has 496 $\times$ 460 arcsec$^2$ and the bulk of the groups emission is
inside a circle with $\sim$180 arcsec of radius.

In each pixel we try a spectral
fit to determine  the temperature and metallicity simultaneously.
We set the minimum count number of 1200 necessary for proceeding with a spectral fit using
\texttt{APEC} plasma model. With this count number we have uncertainties in the temperature map not higher than
5\%, and in the metallicity map the errors are around 7\% (and 12\% beyond $\sim$ 100 kpc).

If we do not have the minimum count number in a pixel,
we try a square region of 3 $\times$ 3 pixels; and if we still do not have the
minimum number of counts, we try a 5 $\times$ 5 pixel region.  If we
still do not have enough counts, the pixel is ignored and we proceed
to the next neighbouring pixel. This is done for all pixels in the
grid.
When we do have enough counts, the best-fit values are attributed to the 
central pixel. 
Since the spectral extraction regions are typically larger
than the pixel map, individual pixel values are not independent.

For our analysis, we produced temperature, metallicity, pressure, and entropy maps.
The pressure and entropy distributions are called pseudo maps to reflect that both pressure and entropy are local
quantities, while the projected emission-weighted temperature and X-ray surface brightness are averaged along the line
of sight. Although surface brightness analysis and temperature maps (used as a tracer of the dynamical history of the system) 
are more commonly used in studies of clusters and
groups of galaxies, pressure, and entropy maps are more straightforward to interpret.  
Pressure and entropy come from their definitions ($nkT$ and $kTn^{-2/3}$, respectively) and thus $P \propto T \times I^{1/2}$
and $S \propto T/I^{1/3}$, where \textit{I} is the intensity (net counts/arcsec$^{2}$).

In the case of a system in hydrostatic equilibrium, pressure fluctuations trace departure from local equilibrium.
Common examples are shocks \citep{mark02,finog04,simi09} and pressure waves \citep{fabian03,schuecker04}.

The entropy map for a system in equilibrium should be symmetrical around the centre and exhibit an increasing
entropy level towards the outskirts. Low-entropy gas displaced from the centre  is either due to instabilities like cold fronts
\citep[e.g.][]{MV07} or to stripping \citep{finog04}. Areas of high-entropy gas are produced by local heating, most likely by AGNs.
Thus,  the projected pseudo-entropy and pressure maps can provide direct information about the 
non-thermal heating processes of the IGM.

\section{X-ray results of the NGC4325 group}
\label{res}

To compute  the gas mass we first converted the surface
brightness distribution into a projected emissivity profile, which was
modelled by a $\beta$-model \citep{CFF78}.  The gas mass is given by
\begin{equation}
\label{Mgas}
M_{\rm gas}(r) = 4~\pi~m_{p}~\mu_{e} \int_{0}^{r_{\Delta}} n_{e}(r) r^{2} dr, 
\end{equation}
and for the $\beta$-model we can write
\begin{equation}
n_{e}=\frac{n_{0}}{[1+(\frac{r}{r_c})^2]^{\frac{3\beta}{2}}},
\end{equation}
where $r_{c}$ and $\beta$ are the characteristic radius and the slope
of the surface brightness profile, $\mu_{e}=1.25$, and $n_{0}$ is the central density
obtained from the normalization parameter from the spectra.

To compute the total mass based on X-ray data, we rely on the
assumption of hydrostatic equilibrium (HE) and spherical symmetry. 
The total mass can be calculated using the deprojected surface brightness
and temperature profiles. The total mass is given by
\begin{equation}
\label{Mtot}
M_{\rm tot} (< r_{\Delta}) = - \frac{k_{b} T r}{G \mu m_{p}}\big(\frac{d \ln \rho}{d \ln r} + \frac{d \ln T}{d \ln r}\big),
\end{equation}
where $r_{\Delta}$ is the radius inside which the mean density is
higher than the critical value by a factor of $\Delta$ (in our case,
$\Delta = 2500$ and $\Delta=500$) , $k_{b}$ is the Boltzman constant, T
the  gas temperature, $m_{p}$ the proton mass, $\mu$ the
molecular weight, and $\rho$ the gas density.

\begin{figure}[ht]
\centering
\includegraphics[scale=0.7]{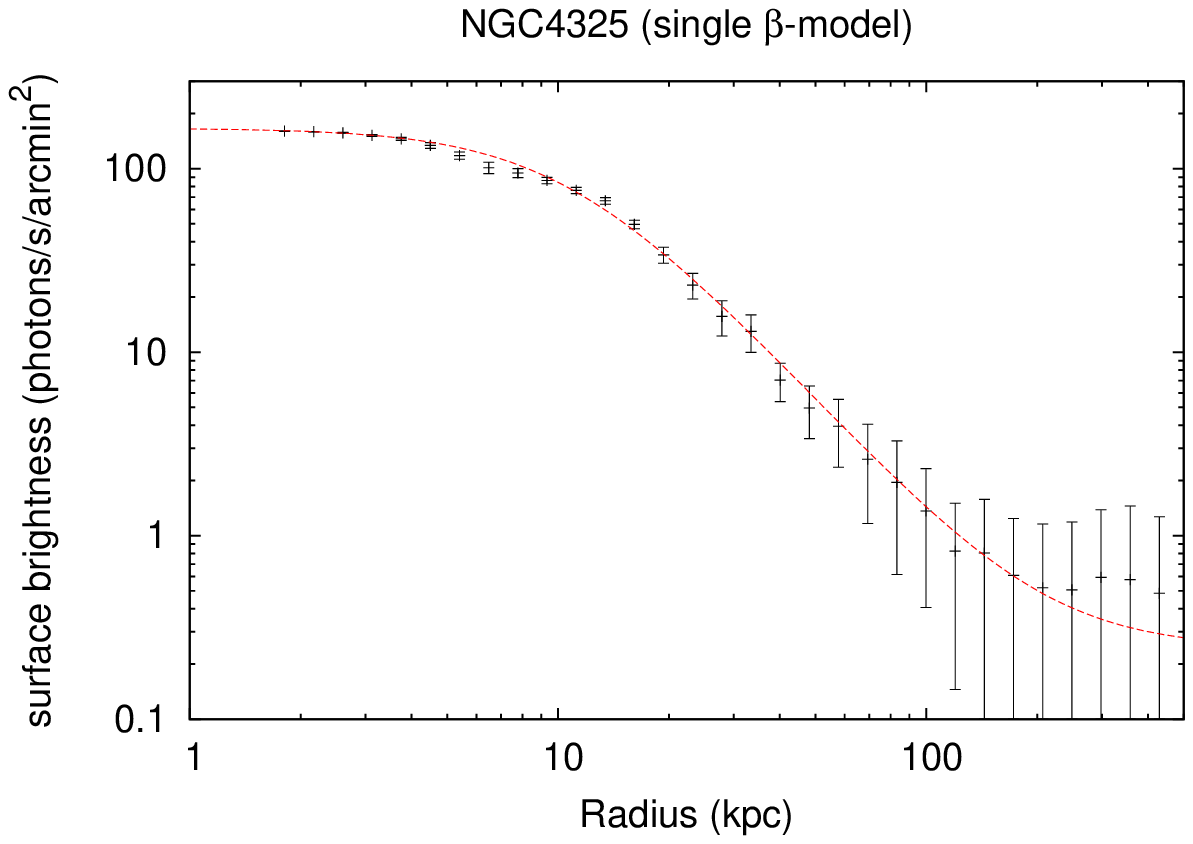}
\includegraphics[scale=0.7]{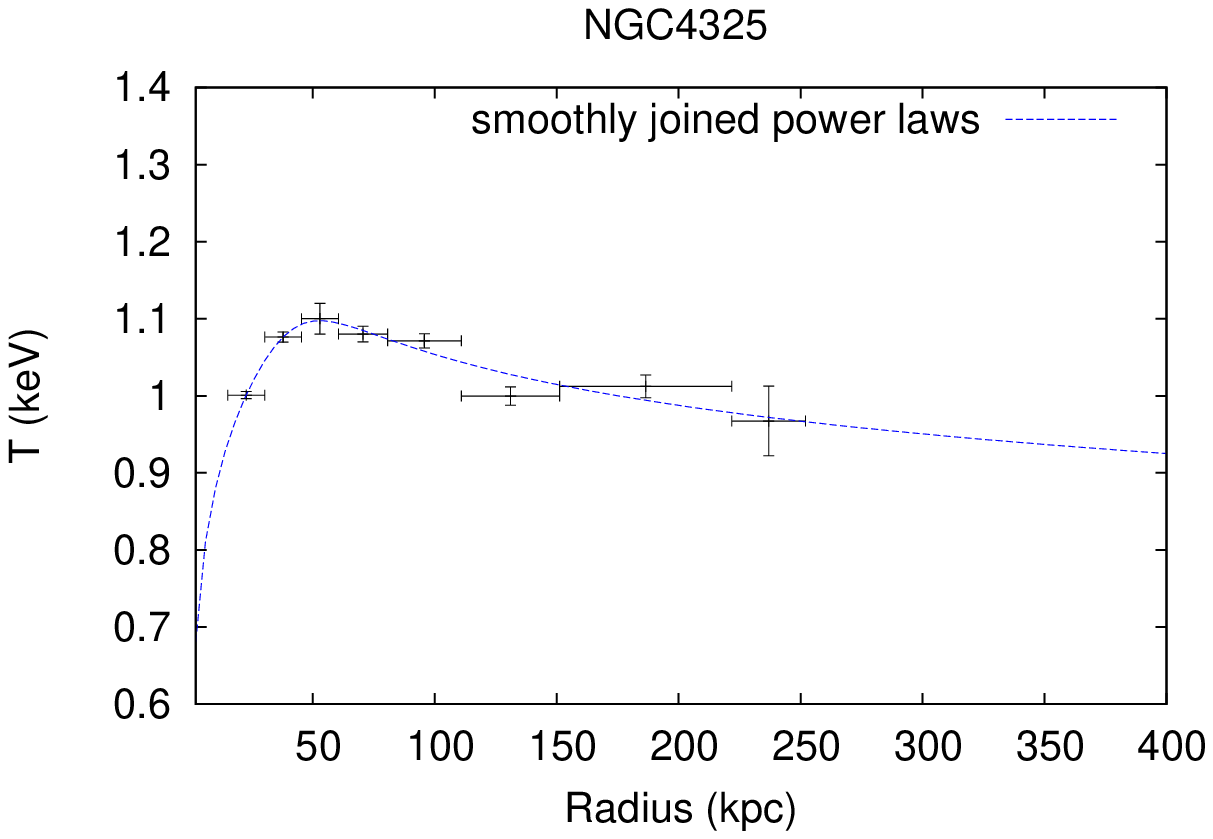}
\caption{X-ray results for the fit of the surface brightness and temperature profiles. 
Upper panel: the black points represent the surface brightness and the red-dashed line represent the single $\beta$-model fit. 
Lower panel: the black point represent the measured temperature and the blue-dashed line represents the smoothly
joined power laws fit.}
\label{fig:profiles}
\end{figure}

We used a single $\beta$-model \citep{CFF78} to describe the surface brightness profile and
 smoothly joined power laws  \citep{Gastaldello07} to fit the temperature of this group. The temperature and
surface brightness fits are shown in Fig.\ref{fig:profiles}. The $r_{2500}$, $r_{500}$,  gas, and total masses  computed 
within these radius are presented in Table~\ref{tab:xray}. It is important to mention that to compute the gas and total mass within
$r_{500}$, we need an extrapolation of the temperature and surface brightness profiles leading to higher uncertainties in the masses.
However, our results agree within 1$\sigma$ with the ones derived by \citet{Gastaldello07} and \citet{lovisari14}, except the gas mass
values computed within $r_{2500}$ that are comparable within 3$\sigma$.

\begin{table*}[ht]
\caption{Mass determination} 
\centering
\begin{tabular}{c | c c c | c c c}
\hline\hline
\noalign{\smallskip}
Group & $r_{2500}$ &  $ M_{\rm gas,2500}$ & $M_{\rm tot,2500}$ & $r_{500}$  & $M_{\rm gas,500}$ &  $M_{\rm tot,500}$  \\

& ($h_{70}^{-1}$ kpc) &  ($10^{12} h_{70}^{-5/2} M_{\odot}$) & ($10^{13} h_{70}^{-1}  M_{\odot}$)  &  ($h_{70}^{-1}$ kpc) &  ($10^{12} h_{70}^{-5/2} M_{\odot}$)& ($10^{13} h_{70}^{-1} M_{\odot}$)  \\
\hline
\noalign{\smallskip}
NGC 4325 & $207^{+6}_{-5}$ & $0.52^{+0.03}_{-0.04}$&  $1.30^{+0.10}_{-0.09}$ & $444^{+12}_{-10}$ &    $1.5^{+0.19}_{-0.20}$ &  $2.57^{+0.19}_{-0.18}$ \\
\noalign{\smallskip}
\hline
\end{tabular}
\label{tab:xray} 
\end{table*}

\begin{figure}[ht!]
\centering
\includegraphics[scale=0.4,angle=90]{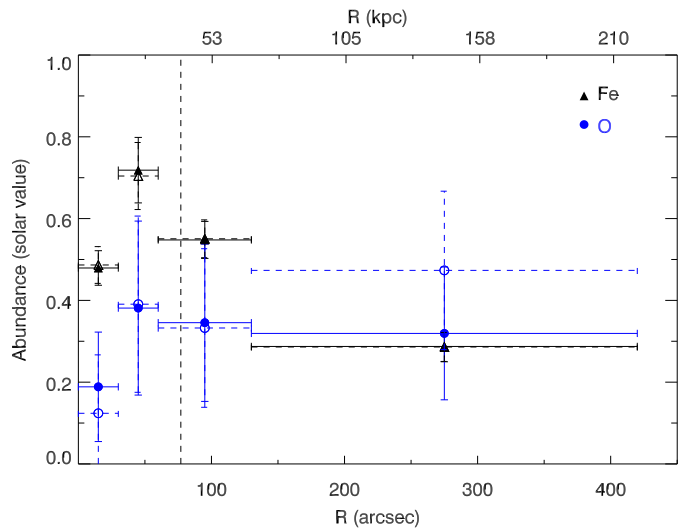}
\caption{Iron and oxygen abundance profiles. The blue dots represent the oxygen abundance with respect to solar values. The black triangles
represent the iron abundance with respect to solar value. The open symbols represent the values computed with $nH$ free to vary.
The vertical dashed line represents the radius within which the high-metallicity structure is enclosed
($r \sim 77$ arcsec).}
\label{fig:FeOprof}
\end{figure}

In Fig.\ref{fig:FeOprof}, we present the iron and oxygen profiles. 
The measure of oxygen abundances might be problematic \citep[although used in abundance analysis as in][]{dePlaa06,werner06,david11,lovisari11,bulbul12}
because the spectrum of the Galactic
warm-hot X-ray emitting gas also contains {O VIII} lines that cannot be separately detected with the spectral resolution of EPIC  \citep[e.g.][]{dePlaa07}
and also because the obtained value is sensitive to the choice of $n_{\rm H}$ value.
Thus, we also left this parameter free to vary (Fig.\ref{fig:FeOprof}), 
showing that our results are robust (all values agree within 1$\sigma$).
Our results show that the oxygen abundance is radially more constant (values around 0.35 solar), while the iron  abundance
profile shows central enhancements, specifically inside $\sim $150 arcsec \citep[in line with results for galaxy clusters presented by e.g.][]{Tamura04,dePlaa07}.

\begin{figure*}[ht]
\centering
\includegraphics[scale=0.4]{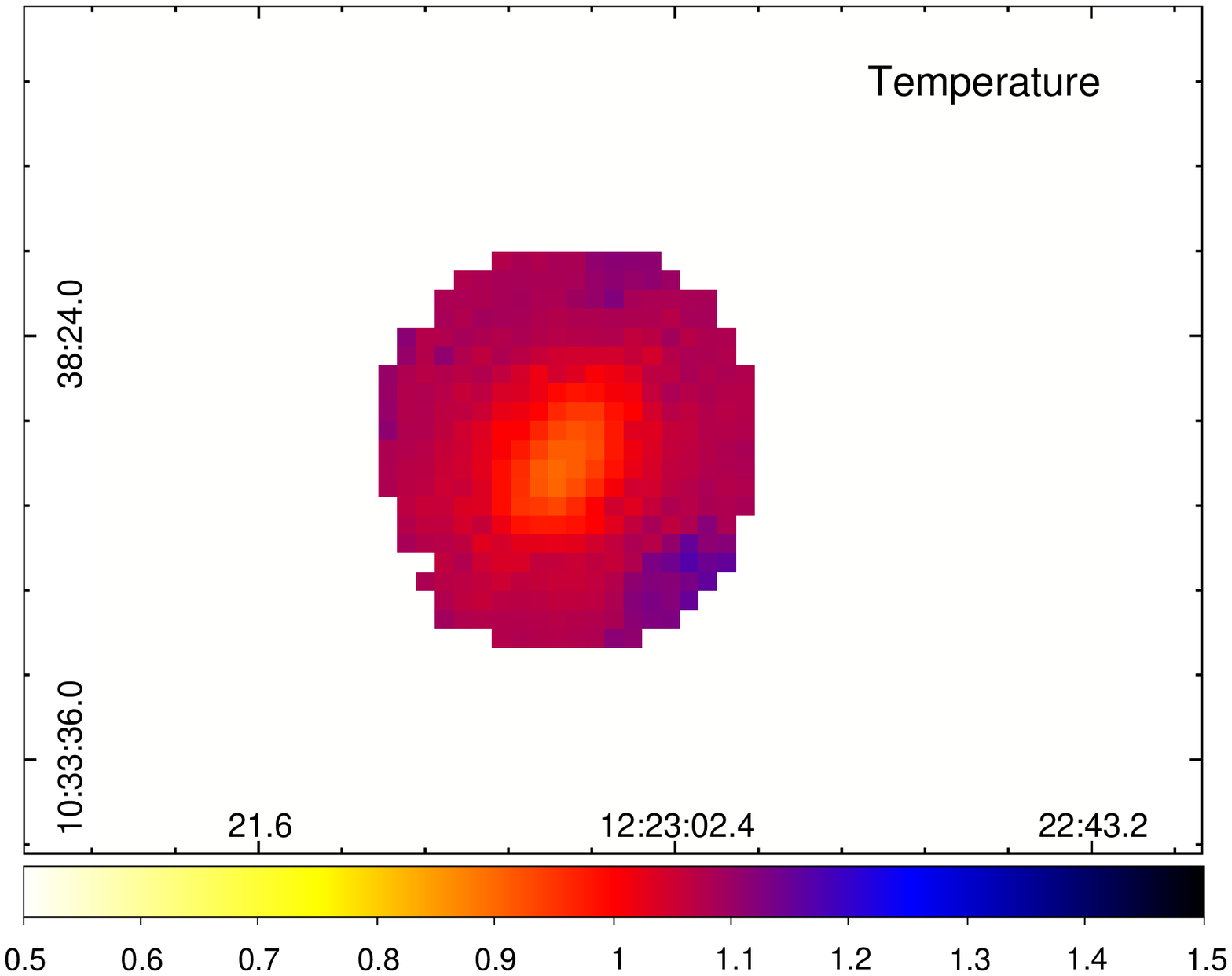}
\includegraphics[scale=0.4]{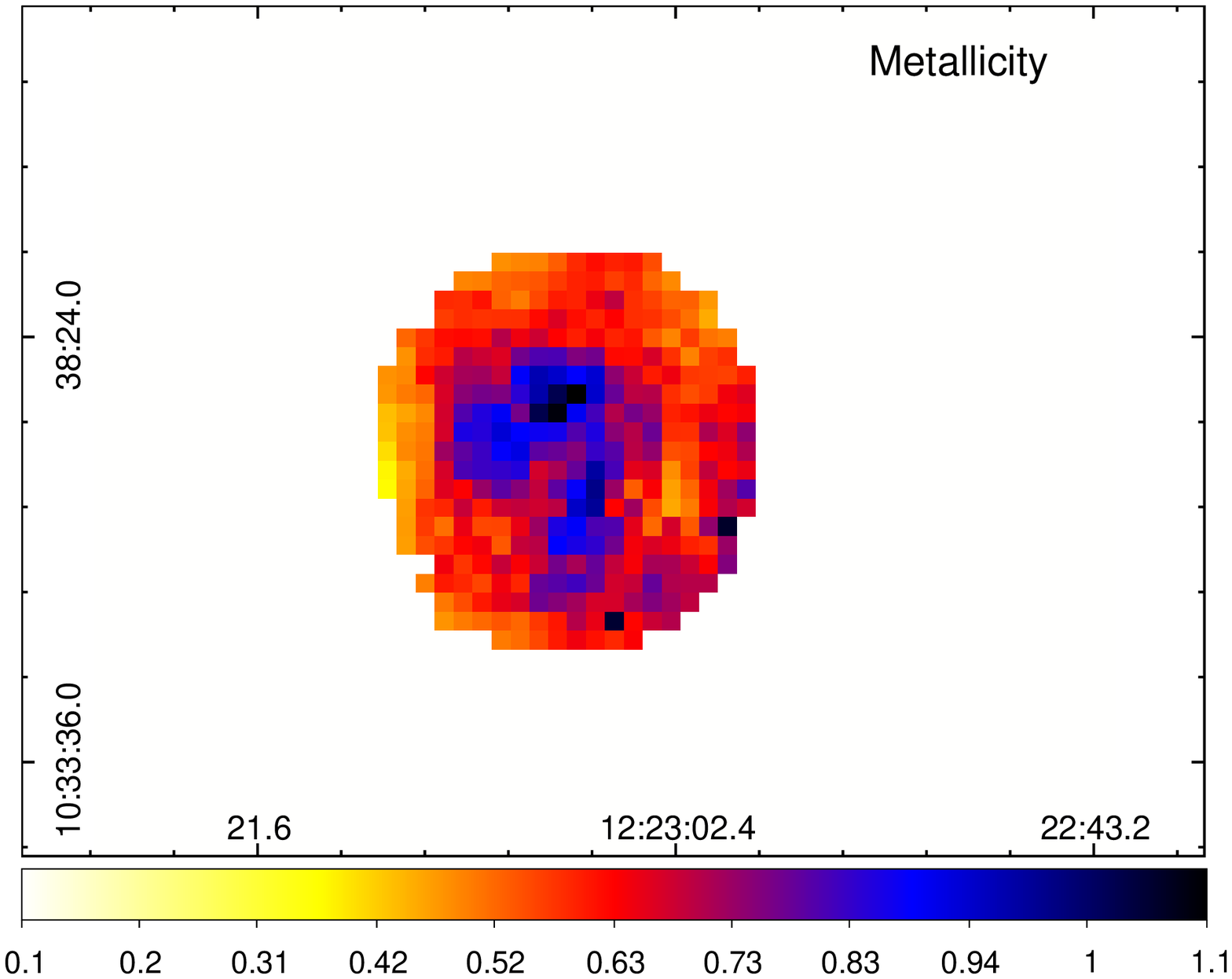}
\includegraphics[scale=0.4]{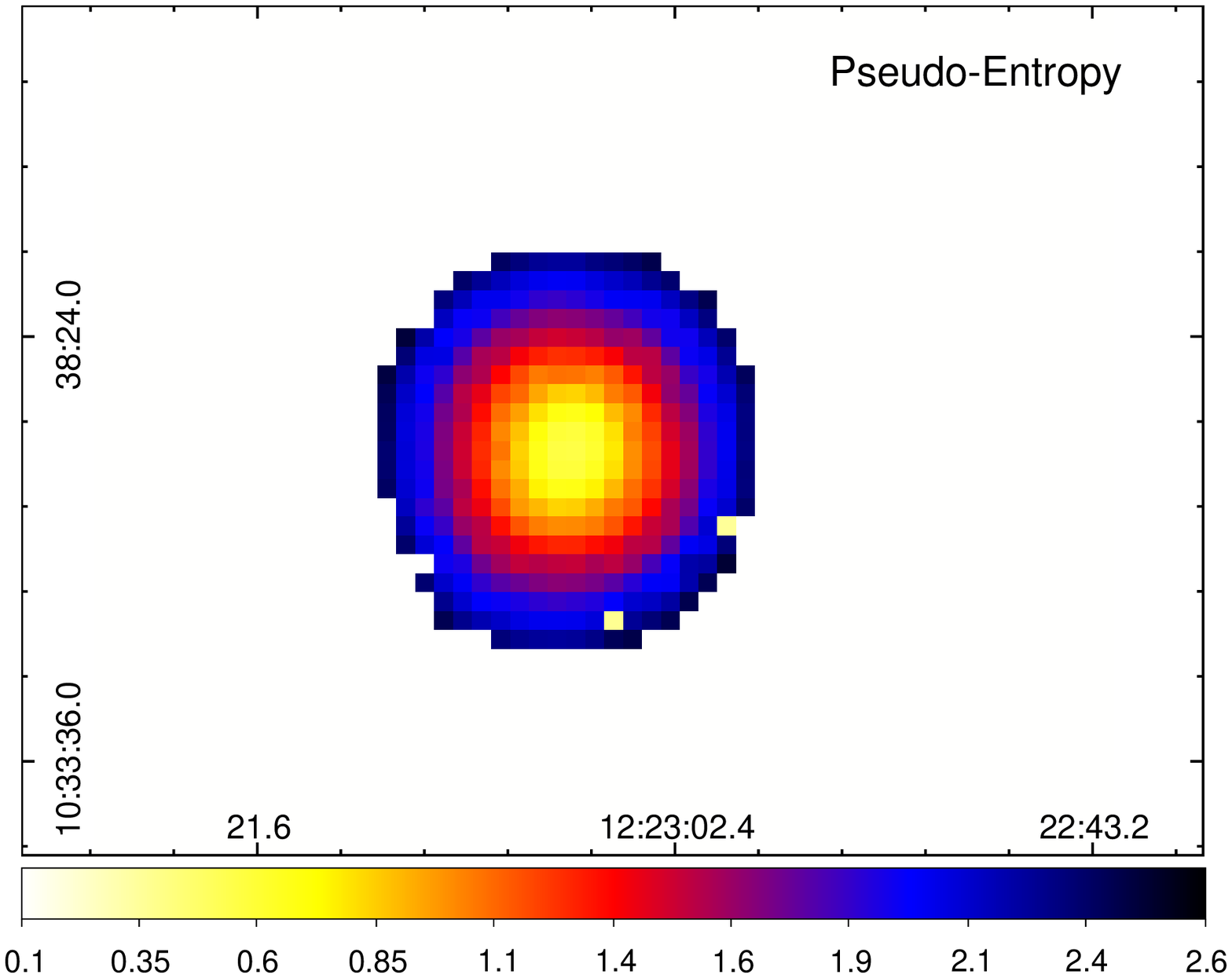}
\includegraphics[scale=0.4]{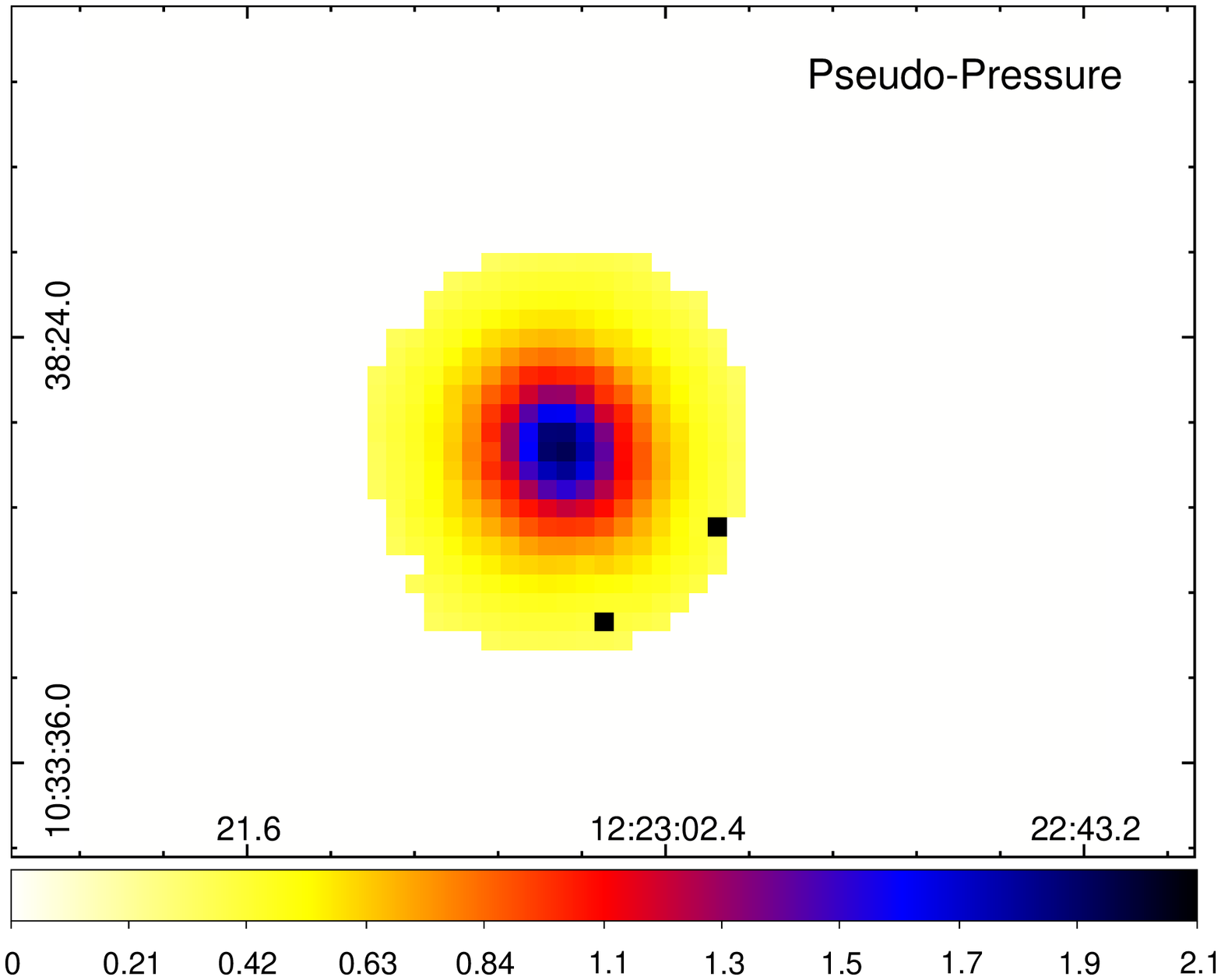}
\caption{Top left: Temperature map. Top  right: metallicity map.
Bottom left: Pseudo-entropy 2D map  derived from $T/I^{1/3}$. Bottom right:  pseudo-pressure 2D map derived from $T \times I^{1/2}$.
Colour bars give values of the quantities. keV for temperature, $Z_{\odot}$ for metallicity, and arbitrary units for pressure and entropy. 
Coordinates are for J2000.0}
\label{fig:maps}
\end{figure*}

For a better analysis, we present in Fig.\ref{fig:maps} the spatial distribution of the temperature, iron abundance, projected pseudo-entropy, and 
projected pseudo-pressure. 
While the temperature map shows no obvious deviations from the almost isothermal behaviour (indicative of no strong cool-core), 
the iron abundance  is not 
spherically distributed. There is an elongated structure in the core of the group in which the metallicity rises from an overall value of  $(0.6 \pm 0.08) Z_{\odot}$ 
to  $(0.9 \pm 0.07) Z_{\odot}$ in this filament.

\begin{figure*}[ht!]
\centering
\includegraphics[scale=0.9]{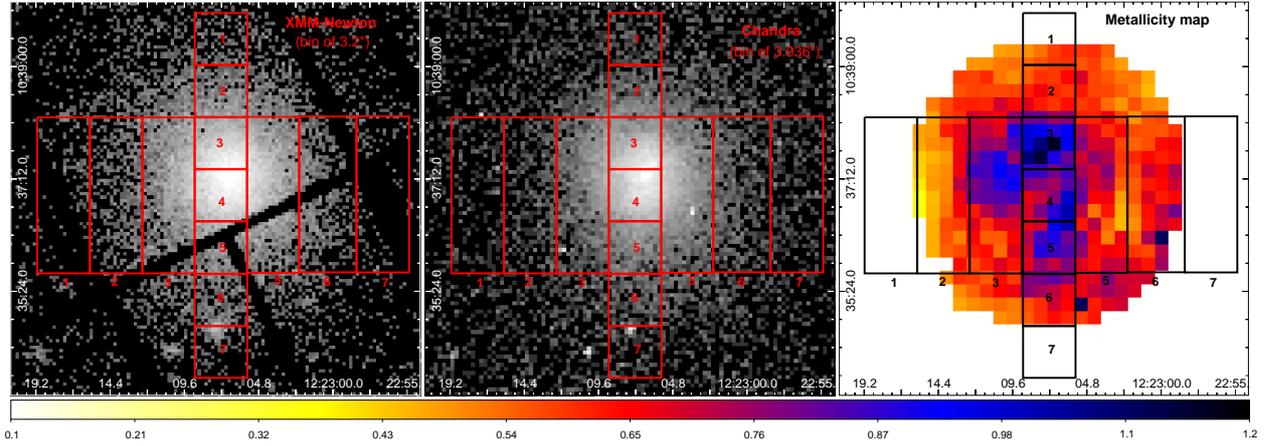}
\caption{XMM-\textit{Newton} image (left panel), \textit{Chandra} image (central panel) and metallicity map (right panel) overlaid with the rectangular regions 
used to examine the variation in metallicity along and across the filament position. 
The colourbar refers to the metallicity map and values are given in $\rm Z_{\odot}$.}
\label{fig:boxes}
\end{figure*}

\begin{figure}[ht]
\centering
\includegraphics[scale=0.4,angle=90]{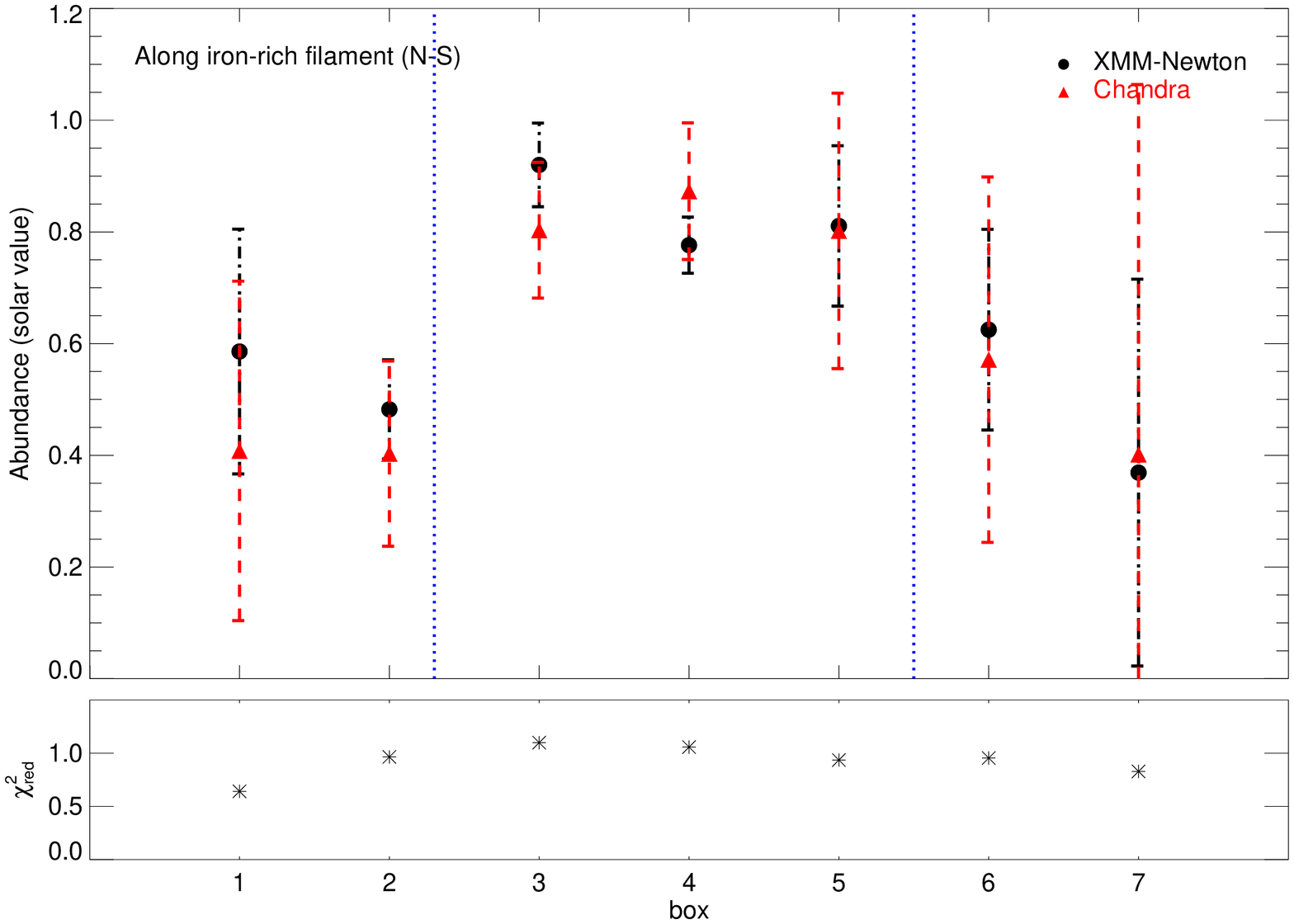}\\
\includegraphics[scale=0.4,angle=90]{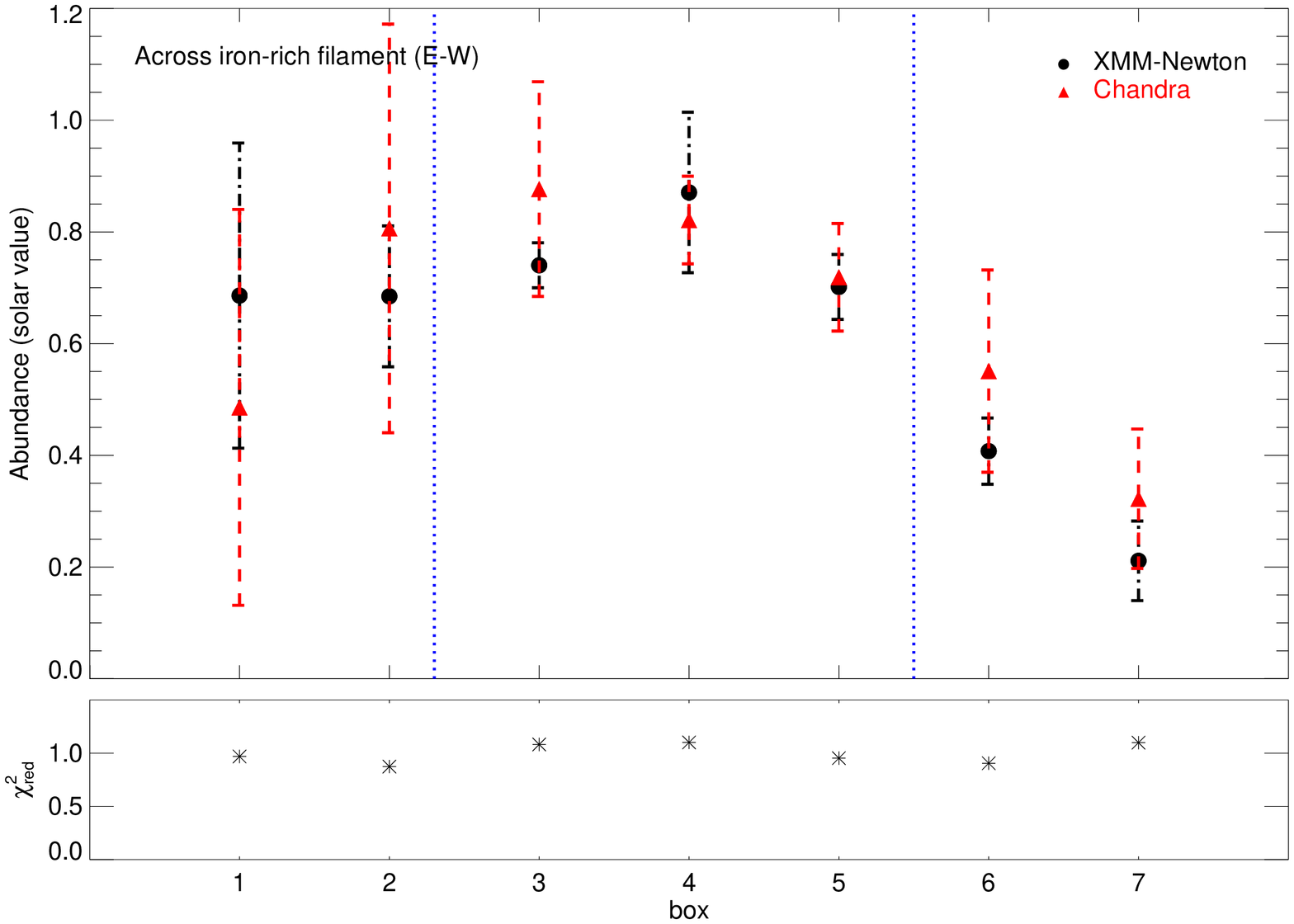}
\caption{Measured metallicity in profiles running north to south along the metal-rich filament (upper panel) and
east to west across the filament. The black dots refer to XMM-\textit{Newton} metal-abundance values, while the red triangles refer to the \textit{Chandra} values.
The  XMM-\textit{Newton} $\chi_{\rm red}^{2}$ of the fit in the corresponding box is plotted below each panel.
The blue lines represent where the filament is located relative to the boxes. 
The numbers on the \textit{x}-axis are the numbers of the boxes in Fig.\ref{fig:boxes}.}
\label{fig:Zprofiles}
\end{figure}

To better examine the elongated filament found in the metallicity map, we placed a number of rectangular regions along and across
this structure, as shown in Fig.˜\ref{fig:boxes}. 
For the comparison, spectra were extracted
from identical regions in the \textit{Chandra} and XMM-\textit{Newton} data.
Because of the low number of \textit{Chandra} source counts in some of
the regions, we decided to use Cash statistics \citep{Cash79}
for the fitting process in \texttt{XSPEC}.
The resulting abundances are shown in Fig.~\ref{fig:Zprofiles} with the XMM-\textit{Newton} $\chi_{\rm red}^{2}$ of the fit in the
corresponding box  plotted below each panel. 
To compute the east-west profile, which uses larger boxes, 
the central region comprises the three small boxes combined. 
There is  good agreement between XMM-\textit{Newton} and \textit{Chandra} values, and
both profiles show a clear enhancement in metallicity in the 
boxes where the filament is located. This suggests that the metal-rich filament is not a product of poor spectral fits but 
is a real structure of higher abundance.

A similar analysis done by \citet{Sullivan11} also showed an elongated feature of roughly solar abundance 
 in the centre of the poor cluster AWM 4. The structures in temperature and abundance found for this poor cluster
 is another indication of the uneven enrichment of the ICM, owing to the jets of the radio source. 
As for NGC 4325, the location of the high-abundance structures suggests that the material has been entrained and is being transported
 out of the galaxy. This result is furthered supported by recent numerical simulations for galaxy groups as shown in\citet[e.g.][]{gaspari11}

The X-ray projected pseudo pressure and projected pseudo-entropy do not show
inhomogeneities related to non-thermal processes.
The projected pseudo-entropy map closely follows the expectations for a system in equilibrium.

\begin{figure*}[ht]
\centering
\includegraphics[scale=0.9]{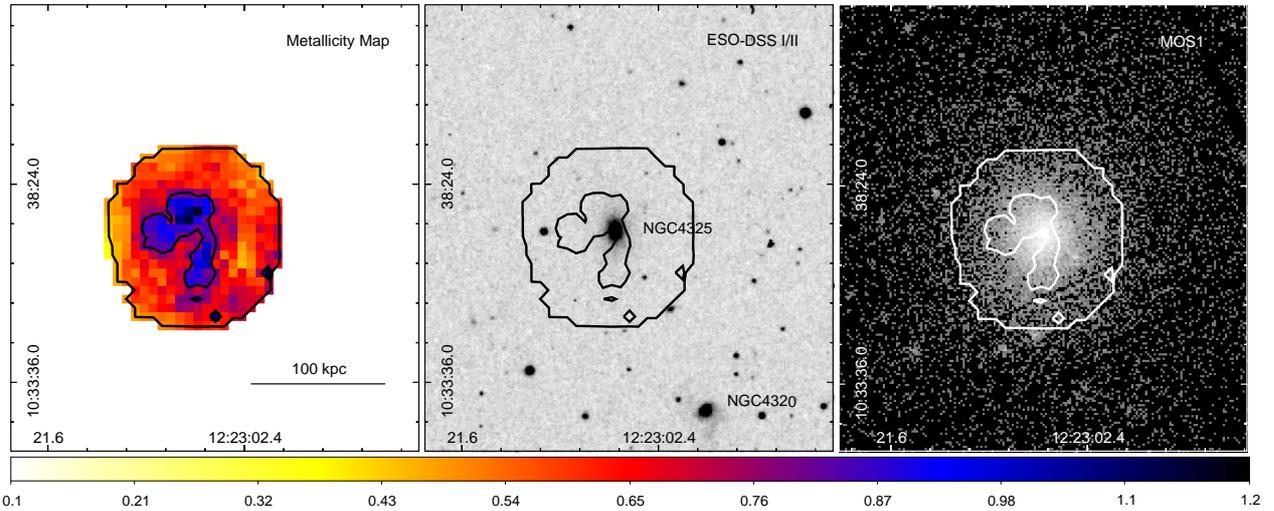}
\caption{Metallicity map (left panel), optical ESO-DSS image (middle panel), and X-ray MOS1 image (right panel) with metal isocontours overlaid. 
The colour bar refers to the metallicity map, and values are given in $\rm Z_{\odot}$.}
\label{fig:ZoptRX}
\end{figure*}

In Fig.~\ref{fig:ZoptRX}, we compare the metallicity map, the optical ESO-DSS I/II image, and the X-ray MOS1 image overlaid with the metal
isocontour. There is a clear spatial correlation between the metal-rich filament and the central galaxy, suggesting a physical connection between the two.
More specifically, because of its elongated morphology, this metal-rich structure could be related to an AGN-outflow.

\section{Stellar population analysis and chemical evolution model of the central galaxy}
\label{cetralgal}

To try to shed some light on which mechanism might be behind the metal enrichment of the IGM, such as an AGN and/or possibly supernovae winds,
 we used stellar population analysis and chemical evolution models. The results are presented in this section.

To perform the stellar population analysis (SPA) we used the optical spectrum of the central galaxy obtained from DR10 Sloan Digital Sky Survey 
\citep[DR10-SDSS,][]{Ahn14}. 
For completeness, we also investigate optical emission line diagnostics.
Then, given the star formation history (SFH) derived by SPA, it is possible to use detailed chemical evolution models to predict the amount of gas 
(and the mass of several specific chemical elements) that a galaxy would have lost by galactic winds during its evolution and analyse its contribution 
to the enrichment of the IGM.

\subsection{Stellar population analysis }
\label{ssp}

The spectral energy distribution (SED) of a galaxy contains an impressive amount of information,
as it is shaped by mass, age, metallicity, dust, and star-formation history of their dominant
stellar populations, as well as the existence (or absence) of an active nucleus. 
One powerful way to extract this information is through the full spectrum fitting technique
\citep[e.g,][]{panter+03,cid+05a,mathis+06,ocvirk+06a,ocvirk+06b, walcher+06, koleva+08}.

The spectral fitting was done using the code \texttt{STARLIGHT}\footnote{The code and its manual can be downloaded 
from http://astro.ufsc.br/starlight/node/1}
\citep{cid+04,cid+05a,mateus+06, asari+07}.
We adopted medium spectral-resolution single stellar population (SSP) models
spanning different ages and metallicities as our spectral base;
i.e. we describe the data in terms of a superposition of multiple bursts of star formation.
The internal kinematics is determined simultaneously with
the population parameters. In addition, we have to take into account the contribution
to the continuum by a possible AGN emission, which would have the form of a
featureless continuum. To represent this featureless
continuum we included a power law in the form of $F_{\nu} \propto \nu^{-0.5}$ in the base set.
Extinction is modelled by \texttt{STARLIGHT} as due to foreground dust and parametrized by the 
V-band extinction (AV). We used the \citet{cardelli+89} extinction law. To accurately model 
the stellar continuum, the emission lines were masked out of the fit.

Because our aim is to shed some light on the nature of the stellar population in the central galaxy
without being biased by a particular choice of models, we opted for three different sets of models: 
\citet{bc03} (hereafter, BC03), \citet{gonzalez+05} (hereafter, GZ05), and \citet{vazdekis+10} (hereafter, VK10). 
For the fit procedure, we selected a range of ages
from the youngest age available for each author  
(1 Myr for GZ05 and BC03 and 63 Myr for VK10) to 15 Gyr, spaced in log(age) = 0.25 dex. For
the metallicities, Z, we adopted three values for each author: below solar, solar, and above solar
(Z = 0.004, 0.020, and Z=0.040 for BC03
and GZ05 and Z = 0.004, 0.019, and 0.030 for VK10). 

To test the sensitivity of the results due to the S/N, we applied the synthesis
to 50 variations in the original spectrum, randomly changing the noise without
changing the actual value of the S/N (which is around 50). This gave us an
error estimation for each age fraction of the population. 
Each variation in the spectrum is created by adding  a random value
($\delta_i$) which can vary between plus and minus each pixel error ($\sigma_i$) 
to the value of each pixel (F$_i$) of the spectrum.

\begin{figure*}[ht]
\centering
\includegraphics[scale=0.45]{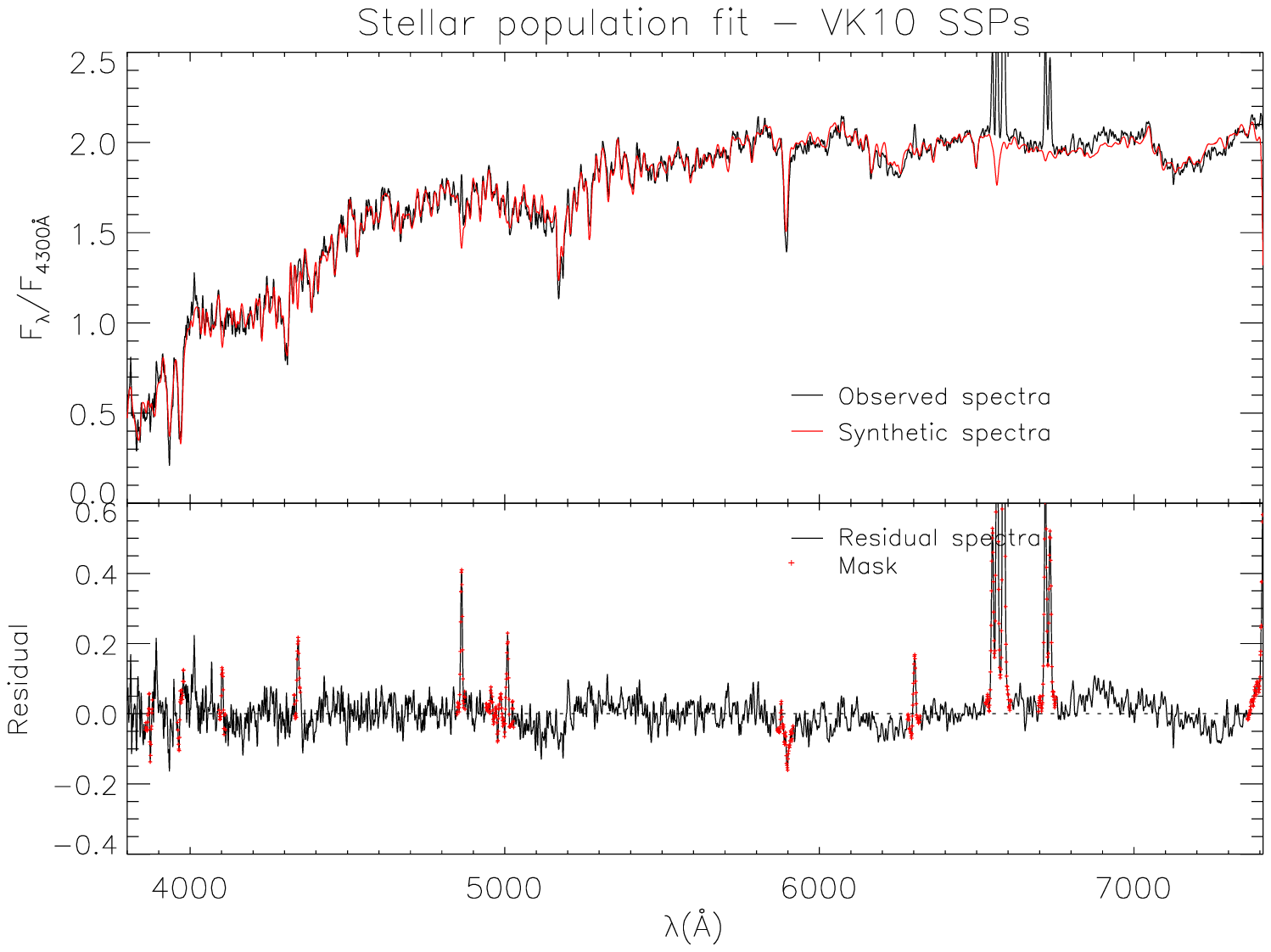}
\includegraphics[scale=0.53]{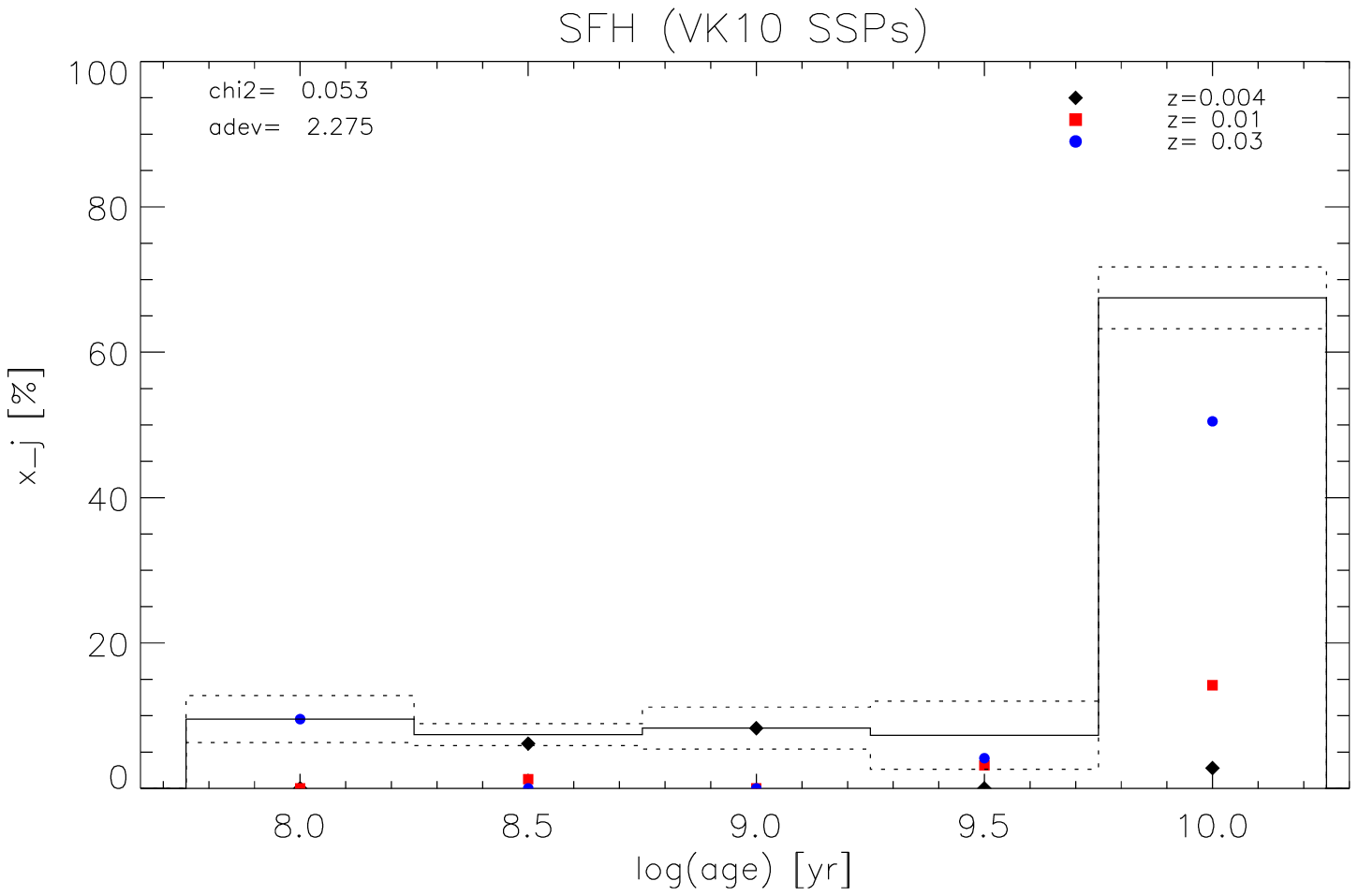}\\
\includegraphics[scale=0.45]{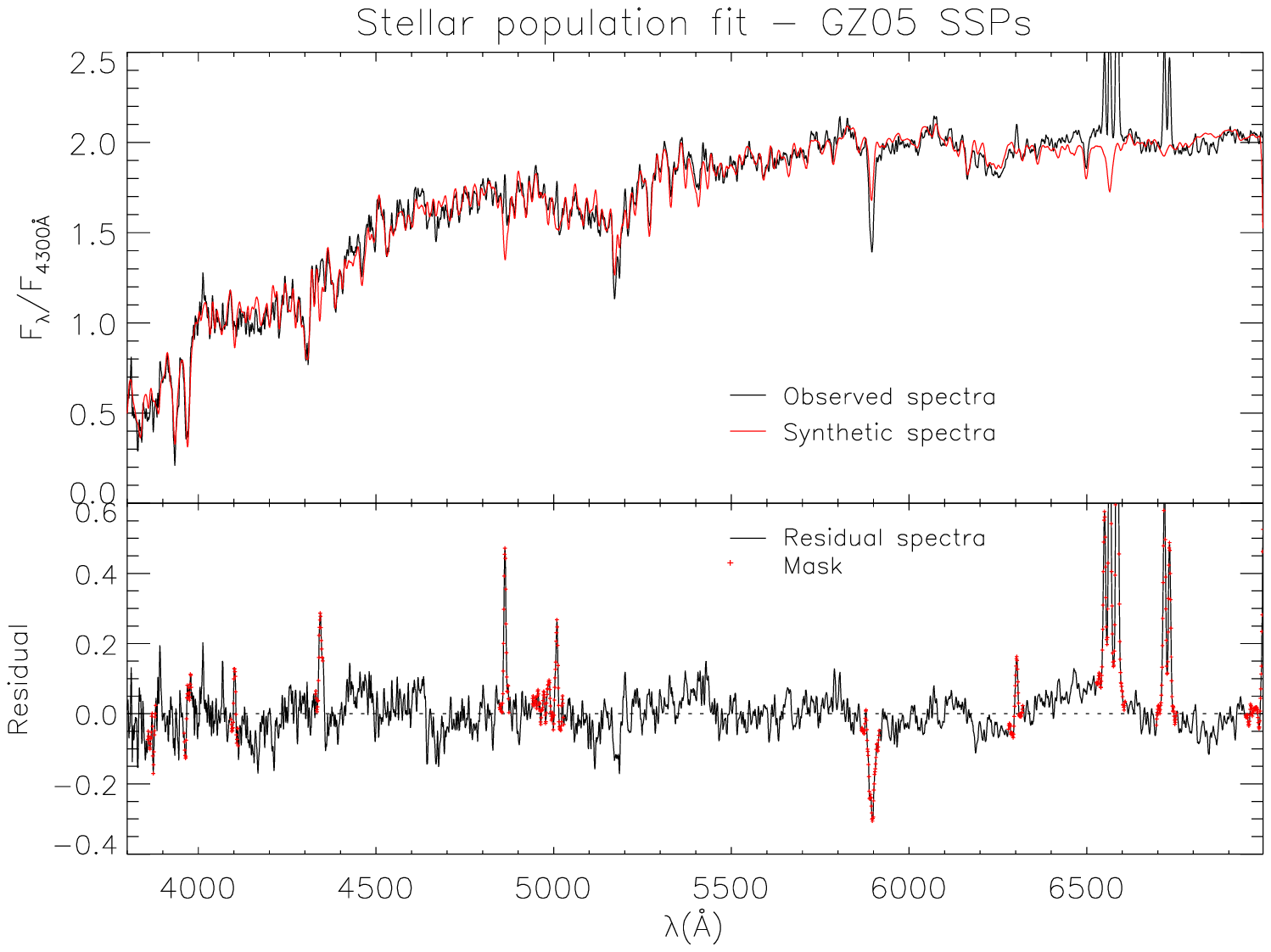}
\includegraphics[scale=0.53]{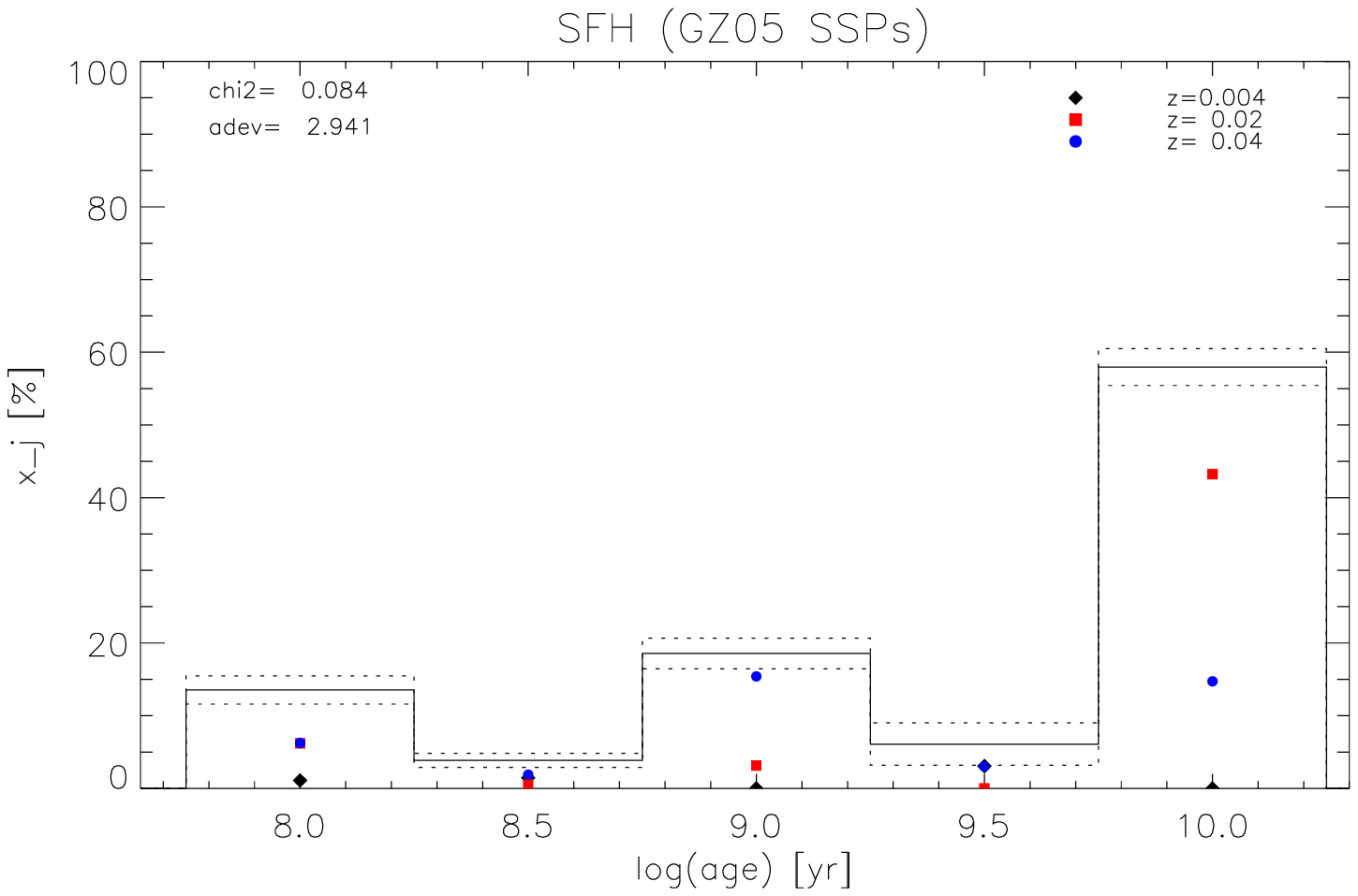}\\
\includegraphics[scale=0.45]{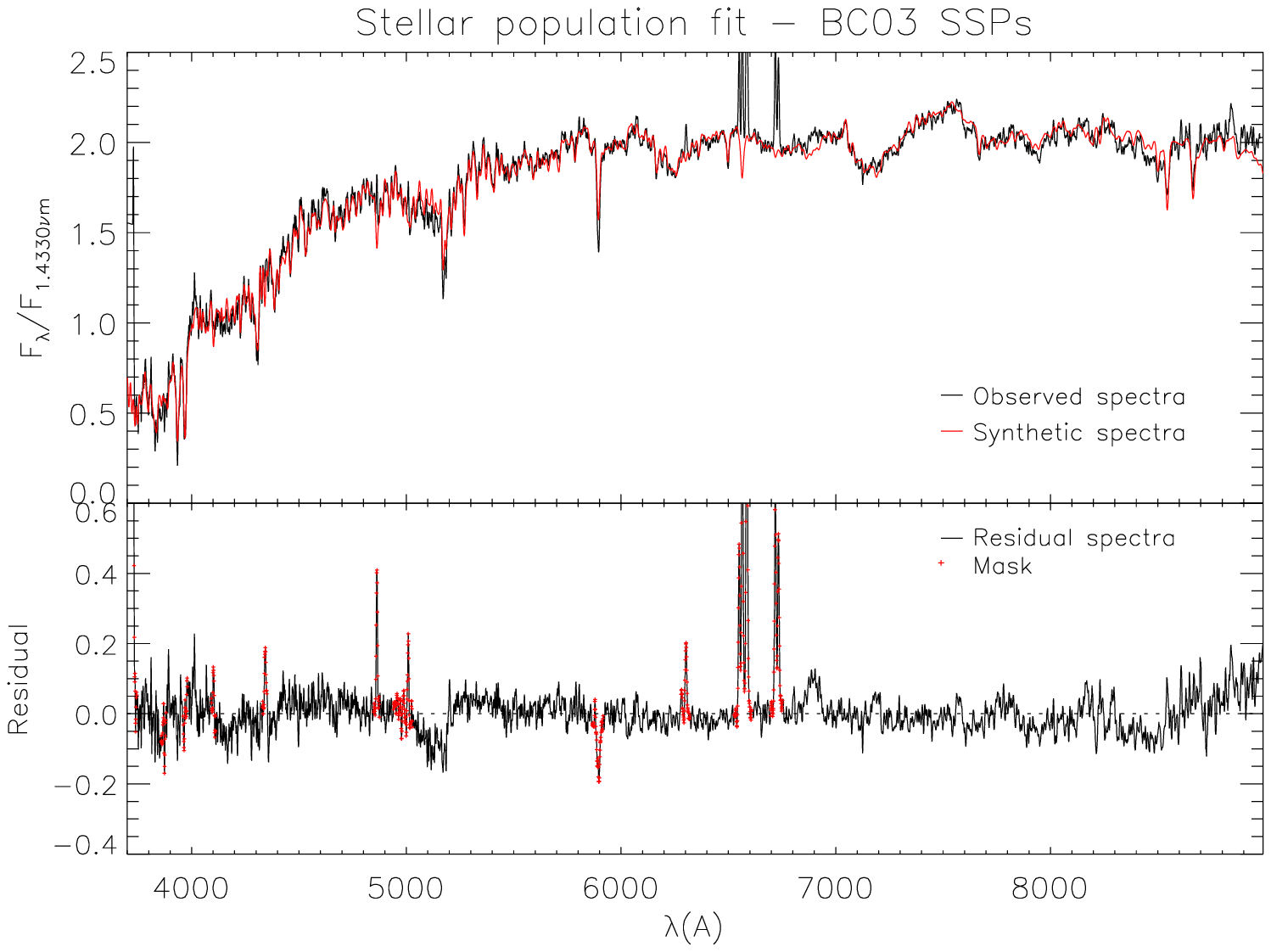}
\includegraphics[scale=0.53]{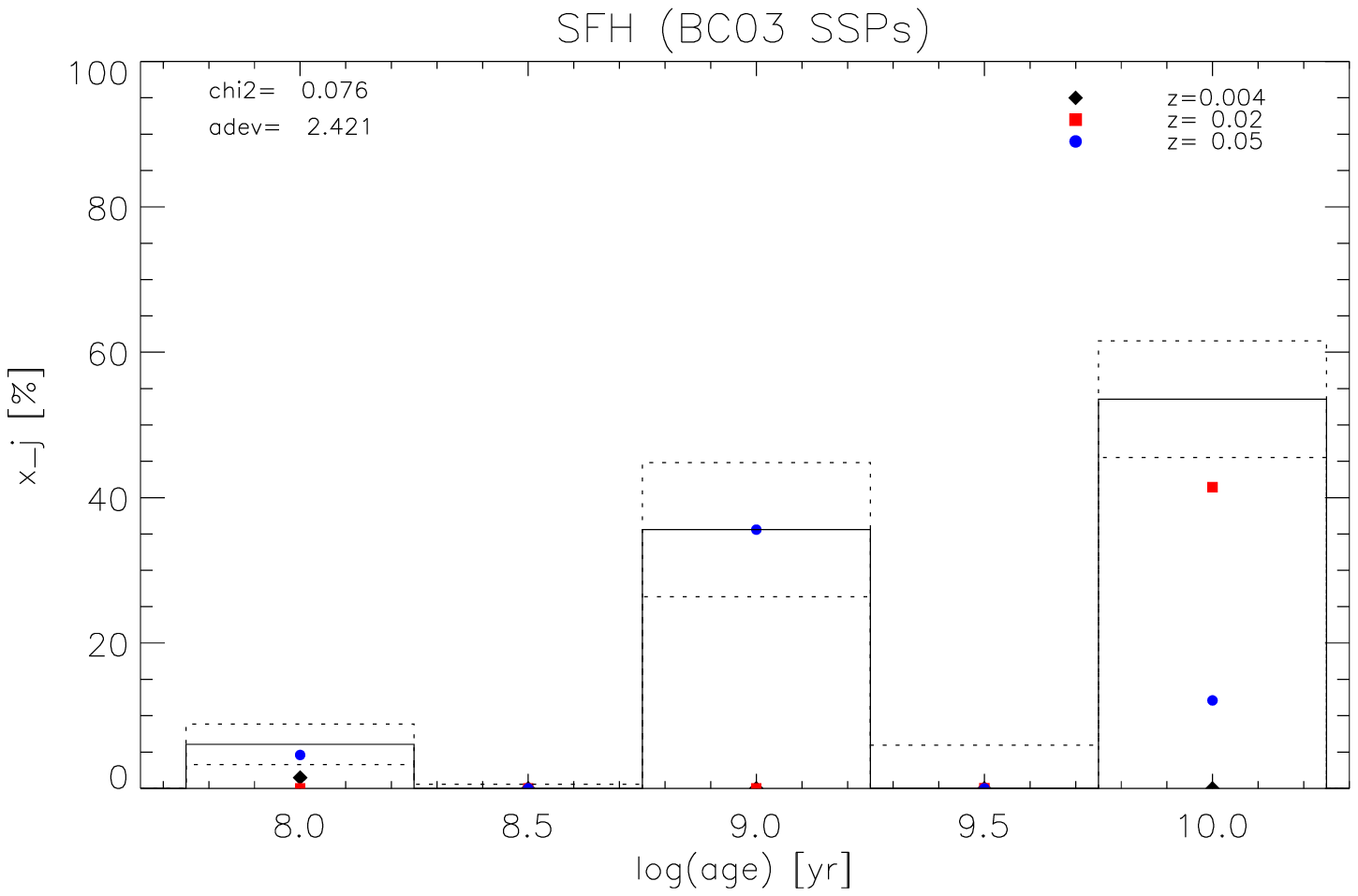}
\caption{Results for the stellar population analysis for the central galaxy NGC4325 using three different base sets: VK10 (upper panels), 
GZ05 (middle panels) and BC03 (lower panels).
\textit{Left column:} The top part of each figure shows the original spectrum (in black) and the synthetic spectrum obtained by our stellar
population fitting (in red), normalised at 4300$\rm \AA$. The bottom panel of each figure shows the residual spectrum (observed minus synthetic).
\textit{Right column}:  the flux fraction ($x_j$) - the fraction of light that comes from the stellar population in that bin - as a function of age. 
The symbols represent the fraction for each metallicity and the black
solid line represents the sum for all three metallicities. The dotted black line marks the error
obtained for each age bin, which is the mean standard deviation from the 50 spectra with
perturbated noise.}
\label{fig:ssp}
\end{figure*}

Figure~\ref{fig:ssp} shows the synthesis results. 
In the left-hand column we show the original spectrum of NGC 4325 (top panels, in black) and the synthetic spectrum obtained
by our stellar population fitting (top panels, in red), normalized at 4300\AA. The bottom panels in each of these plots show the
residual spectrum (observed minus synthetic).
The quality of the fits are measured by ${\chi_\lambda}^2$ and $adev$, as defined 
in \citet{cid+04}. As can be seen from the values presented in the right-hand column of the figure, the fits are very
good for all SSPs models.

From the stellar population synthesis, we can obtain the SFH of the galaxy. 
A description of the SFH in terms of 0.25 dex
age bins is too detailed, given the effects of noise and intrinsic degeneracies
of the synthesis process. A coarser but more robust description of the SFH 
requires further binning of the
age distribution. Because of that we binned the results in five age bins: log(age) 
7.75 $\leq$ log(age) $<$ 8.25 dex,
8.25$\leq$ log(age) $<$ 8.75 dex, 
8.75 $\leq$ log(age) $<$ 9.25 dex,
9.25 $\leq$ log(age) $<$ 9.75 dex,
and log(age) $\geq$ 9.75 dex.
This is shown in the right-hand column of Fig.~\ref{fig:ssp}. The plots show the flux fraction ($x_j$) - 
the fraction of light that comes from the stellar population in that bin - as a function of age. 
The symbols represent the fraction for each metallicity, and the black
solid line represents the sum for all three metallicities. The dotted black line marks the error
obtained for each age bin, which is the mean standard deviation from the 50 spectra with
perturbated noise.

These results show that the galaxy is dominated by an old stellar population, with 55\% to 70\% of the continuum light 
(depending on the base set models) coming from a population older than 5 Gyr.
However, there seems to have been two more recent bursts of star formation, seen
as two peaks in the age distribution (clear in BC03 and GZ05, less obvious with VK10).

There is no evidence that an AGN  contribute to the continuum. For GZ05 and VK10, a
0\% contribution was found by the synthesis. For BC03, \texttt{STARLIGHT} found a small contribution of
5\%, but this can just be compensation for the lack of blue stars for younger
ages in the stellar library they used. This result reinforces the idea that there is no recent AGN activity in this galaxy.
Since emission lines are observed, another mechanism responsible for the gas ionization might be young stars or supernova
shocks.

\subsection{Emission line diagnostics}

Although the NGC~4325  continuum is dominated by the stellar population, 
emission lines are clearly visible in the optical spectrum, as mentioned
in the previous section. The relative intensity of these emission lines can be used
as an ionisation mechanism diagnostic, through diagnostic diagrams like the well known
BPT diagrams \citep{bpt}. 

The idea behind these diagrams is that nebulae photoionized by hot, young stars
can be distinguished from those photoionized by a harder radiation field, such as that of an AGN. 
The BPT diagrams were studied in numerous 
works, and the dividing lines have been developed and adapted as a function of the 
ionization models and/or observations available 
\citep[e.g.][]{veilleux+87,osterbrock89,kewley+01,kauffmann+03,kewley+06,stasinska+06,kewley+13a,kewley+13b}.
For example, for a survey of nearby emission line galaxies \citet{ho+97} separated HII nuclei from AGNs and Liner-like
with the following criteria:
 
\begin{itemize}

\item HII nuclei: [OIII]$\lambda$5007/H$\beta$ - any value; [OI]$\lambda$6300/H$\alpha$ $<$0.08; 
[NII]$\lambda$6584/H$\alpha$ $<$ 0.6 and [SII]$\lambda\lambda$6717,6731/H$\alpha$ $<$ 0.4.

\item Seyfert nuclei:[OIII]$\lambda$5007/H$\beta$ $\geq$ 3; [OI]$\lambda$6300/H$\alpha$ $\geq$0.08; 
[NII]$\lambda$6584/H$\alpha$ $\geq$ 0.6 and [SII]$\lambda\lambda$6717,6731/H$\alpha$ $\geq$ 0.4.

\item Liners: [OIII]$\lambda$5007/H$\beta$ $<$ 3; [OI]$\lambda$6300/H$\alpha$ $\geq$0.17; 
[NII]$\lambda$6584/H$\alpha$ $\geq$ 0.6 and [SII]$\lambda\lambda$6717,6731/H$\alpha$ $\geq$ 0.4.

\end{itemize}

After the subtracting the stellar continuum of NGC~4325, the emission line fluxes can be properly measured. The values of the
key diagnostic emission line fluxes are presented in Table~\ref{emlines}. From these values we obtain the ratios
[OIII]$\lambda$5007/H$\beta$ = 0.61, [OI]$\lambda$6300/H$\alpha$ = 0.24,  
[NII]$\lambda$6584/H$\alpha$ = 1.44 and [SII]$\lambda\lambda$6717,6731/H$\alpha$ =1.17. These values place NGC~4325 not only 
in the LINER regime, but also in the very extreme locus of this region: very strong low-ionization lines and very weak high ionization
lines compared to the hydrogen lines. In the absence of a detectable AGN, it is very likely that the mechanism behind 
the line ionization in this galaxy is shocks of stellar origin, probably from supernova events.

\begin{table}
\centering
\caption{Diagnostic emission line fluxes.}
\begin{tabular}{@{}cc@{}}
\hline
Line &     Flux ($10^{-15} \rm erg/cm^2 /s$)  \\  
\hline
H$\beta$ 4861 &  1.31 $\pm$ 0.04 \\
$\rm [OIII]5007$    &  0.79 $\pm$ 0.07 \\
$\rm [OI]6300 $     &  0.99 $\pm$ 0.05 \\
H$\alpha$6563 &  4.18 $\pm$ 0.09 \\
$\rm [NII]6583$     &  6.00 $\pm$ 0.14 \\
$\rm [SII]6717 $    &  2.74 $\pm$ 0.03 \\
$\rm [SII]6731$     &  2.14 $\pm$ 0.03 \\ 
\hline
\end{tabular}
\label{emlines}
\end{table}

\subsection{Chemical evolution models}
\label{evmodel}

As stars are formed and evolve, energy and chemical elements are released into the interstellar medium (ISM). 
The energy from stars and supernova explosions is accumulated until the kinetic energy of the gas is equal to 
or larger than the potential energy of the galaxy. 
At that point, a galactic wind occurs, removing  a fraction of the gas enriched by fresh elements  from the system \citep[see][]{LM07}. 
The removed material will contribute to the chemical enrichment of the IGM. The epoch when the wind starts, its chemical composition, and 
intensity all depend indirectly on the star formation rate (SFR) and can be inferred by solving the basic equations of chemical evolution 
\citep{Tinsley80,Matteucci98}.

\begin{figure}[ht]
\centering
\includegraphics[scale=0.9]{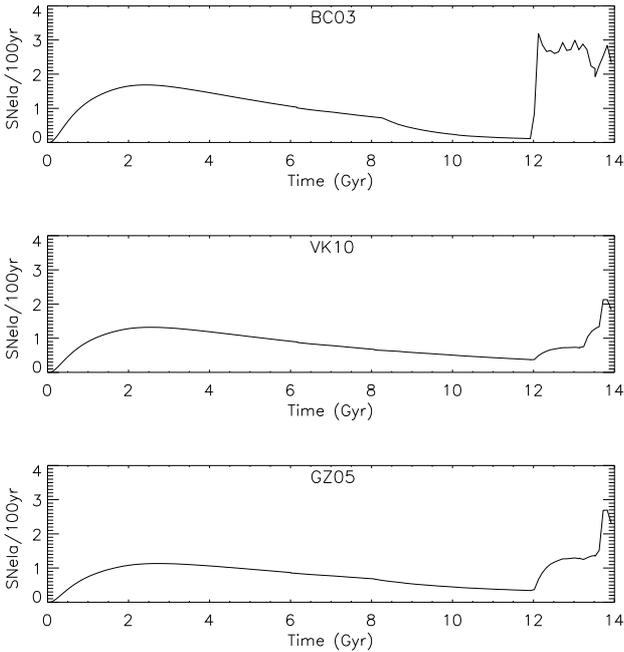}
\caption{Predicted SNe Ia rate as a function of galactic age. The x-axis refers to the $time$ elapsed since the formation of the galaxy, so T = 0 Gyr indicates the initial 
stage and T = 14 Gyr the present time, as opposed to the x-axis of the panels on the right column of Fig.~\ref{fig:ssp}, which represents the $age$ 
of the stellar population that contributes to the spectrum  (so T=0 Gyr represents the youngest stellar and T=14 Gyr the stellar population formed
in the initial stage of the galaxy).}
\label{fig:snia}
\end{figure}

\begin{figure}[ht!]
\centering
\includegraphics[scale=0.9]{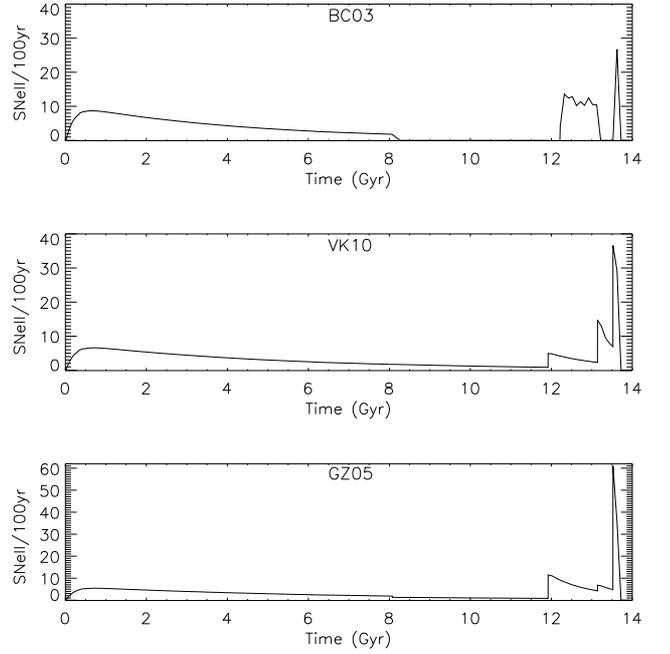}
\caption{Predicted SNe II rate as a function of galactic age. The x-axis refers to the $time$ elapsed since the formation of the galaxy, so T = 0 Gyr indicates the initial 
stage and T = 14 Gyr the present time, as opposed to the x-axis of the panel in the right column of Fig.~\ref{fig:ssp}.}
\label{fig:snii}
\end{figure}


We adopted the code of \citet[][hereafter LM03 and LM04]{LM03,LM04} to simulate the chemical evolution of the galaxy NGC4325. It assumes   
nucleosynthesis and takes the role played by stars of all mass ranges and by supernovae (SNe) of different types (II, Ia) into account 
for the energetics and for the 
chemical enrichment of the galaxy. This allows us to follow the evolution of several chemical elements (H, D, He, C, N, O, Mg, Si, S, Ca and Fe) in detail inside 
and outside the galaxy. The model was adopted for NGC4325, following the formulations of 
\citet[][hereafter PM02]{PM02} and \citet[][hereafter PM04 and PM06]{PM04, PM06}. 

Two very important features that control the occurance of the winds 
(and consequently the amount of material deposited in the IGM) are the binding energy of the galaxy and the kinetic energy of the gas. 
The energy of the gas is defined by the SNe rate \citep[and SFR - for details see][]{Matteucci04} and by the efficiency within which the thermal energy 
is converted into kinetic energy. 
The SFR is given by the SFH history inferred in the previous section (Sect. \ref{ssp}), whereas the conversion of energy is the same as adopted in 
\citet{PM02} and \citet{PM04}.
The binding energy of the galaxy, on the other hand, is mainly controlled by the distribution of dark matter. Following PM04, we use the profile of  
\citet{BSS92}
for the dark matter distribution. The mass of dark matter is ten times higher than the maximum luminous matter, and the radius of the dark matter halo is 
ten times the effective radius of the galaxy (R$_{\rm dark}$ = 10 R$_{\rm eff}$;  PM04). Besides that, we also adopted, as in PM02 and PM04, an infall 
timescale of 0.2 Gyr and a \citet{Salpeter55} initial mass function (IMF).
Small perturbations in the values of the mass and radius of dark matter will not significantly change the results regarding the amount of mass lost by the galaxy. These two parameters mostly influence  the time at which the galactic wind starts: a higher mass of dark matter would delay the occurrence of the wind, whereas a larger radius will anticipate the wind.


\begin{table*}[ht!]
\centering
\caption[]{Model parameters for the NGC4235 galaxy. $M_{\rm lum}$ is the baryonic (luminous) mass of the galaxy, 
SF refers to the epoch (galactic age) and duration of the star formation episodes, 
$R_{\rm eff}$ is the effective radius (assumed to be the DR10-SDSS petroR90, which is
 the radius containing 90\% of Petrosian flux), and IMF is the initial mass function.}
\begin{tabular}{ccccccccc}
\noalign{\smallskip}
\hline
\hline
\noalign{\smallskip}
model &$M_{\rm lum}$  &SF${_1}$ &SF${_2}$ &SF${_3}$ &   SF${_4}$ &SF${_5}$ &$R_{\rm eff}$ &IMF\\
 &$(M_{\odot})$ & (Gyr) & (Gyr) & (Gyr)& (Gyr)& (Gyr)& ($h_{70}^{-1}$ kpc)  & \\
\noalign{\smallskip}
\hline
\noalign{\smallskip}
VK10 &$6 \times 10^{10}$  &0 - 8.08 &8.08 - 11.9 &11.9 - 13.1 &13.1 - 13.5 &13.5 - 13.6 &14.0 &Salpeter\\
GZ05 &$6 \times 10^{10}$  &0 - 8.08 &8.08 - 11.9 &11.9 - 13.1 &13.1 - 13.5 &13.5 - 13.6 &14.0 &Salpeter\\
BC03 &$6 \times 10^{10}$  &0 - 8.08 &11.9 - 13.1 &13.5 - 13.6 &-- &-- &14.0 &Salpeter\\
\hline
\end{tabular}
\label{tabevmod1}
\end{table*}


The NCG 4325 model is specified by the star formation history inferred in Sect.\ref{ssp} and by observational features of the galaxy, 
such as the total luminous mass and 
the effective radius (see Table \ref{tabevmod1} for more details). 
 The adopted scenarios for the SF in
 this galaxy consist of three or five episodes of activity, depending on the SFH adopted (BC03 - three episodes; 
VK10 and GZ05 - five episodes). Results of the stellar population synthesis indicate that the majority of the stellar 
light in all models (56.25$\%$ in BC03, 67.0$\%$ in VK10, and 58.0$\%$ in GZ05) comes from the first episode of SF,
 $\sim$ 20.0 - 35.0$\%$ from intermediate epochs, and the remaining  $\sim$ 10.0 - 15.0$\%$ of the stellar light from the
 SF in the last hundred Myr. To reproduce these results, the adopted efficiencies of the SF should differ among the episodes,
 yielding different star formation rates. In all three models, the SFR is low at the beginning of the evolution of the galaxy 
and decreases owing the consumption of the interstellar gas until intermediate epochs. In the last two Gyrs it increases again.
 With this choice of SF, we were able to exactly reproduce the SFHs given before (and shown in Fig.\ref{fig:ssp}).


\begin{table} 
\caption[]{Predicted amount of mass released by NGC4235 galaxy up to the present time and the present stellar mass. $M_{\rm wind}$ is total mass of gas 
lost through wind, $M_{\rm Fe}$ is the mass of iron, $M_{\rm O}$ is the mass of oxygen, and $M_{\rm star}$ is the present-day stellar mass.}
\centering
\begin{tabular}{ccccc}
\hline
\hline
Model & $M_{\rm wind} $  &$M_{\rm Fe} $  &$M_{\rm O} $  &$M_{\rm star} $\\
 & $(M_{\odot})$ & $(M_{\odot})$ & $(M_{\odot})$ & $(M_{\odot})$ \\
\noalign{\smallskip}
\hline
\noalign{\smallskip}
VK10 &$4.36 \times 10^{10}$  &$0.91 \times 10^{8}$ &$4.37 \times 10^{8}$ &$5.84 \times 10^{10}$\\
GZ05 &$5.78 \times 10^{10}$  &$0.79 \times 10^{8}$ &$3.84 \times 10^{8}$ &$4.52 \times 10^{10}$\\
BC03 &$4.53 \times 10^{10}$  &$1.01 \times 10^{8}$ &$4.65 \times 10^{8}$ &$6.89 \times 10^{10}$\\
\hline
\end{tabular}
\label{tab:evmod2}
\end{table}

In all models, a galactic wind is developed at early stages of galactic evolution, removing a fraction of the ISM of the galaxy and injecting the 
enriched gas into the IGM. In NGC 4325, the wind starts at a galactic age between 0.22 and 0.40 Gyr (depending on the model), but unlike PM models the 
SF is not halted after the onset of the wind since the stellar population analysis indicate a very prolonged initial episode of SF followed by other ones until very 
recent times. As soon as the wind starts, the fraction of O that is injected into the IGM is higher than Fe due to the small timescale for the production of O by 
massive stars and posterior release into the ISM by SNe II. 
As the evolution proceeds, however, the amount of Fe in the ISM and, consequently in the wind, gets higher 
due to the increase in the number of supernova type Ia (SNe Ia). Since this type of SN originates in an intermediate mass star in a binary system, 
its timescale is longer. This gives rise to a SNe Ia rate that does not goes down to zero even when the SF is not active (see Fig.\ref{fig:snia}). 

By solving the equations 
given in PM02 (see their section 3), we were able to calculate the total mass of gas, the mass of Fe, and the mass of O released by the galaxy in the IGM through 
galactic winds. As shown in Table \ref{tab:evmod2}, during the evolution of the galaxy, the total amount of O that is injected into the IGM by the galaxy NGC4325 is higher 
than the mass 
of Fe by a factor between 4 and 5. Even though the SNe Ia rate does not go to zero, unlike SNe II rate, which follows the SFH and is negligible in periods of quiescence 
(as shown in Fig.\ref{fig:snii}), the total amount of O present in the winds is higher than the amount of Fe because the rate of SNe II (main producer of O) is much higher than the 
rate of SNe Ia (main producer of Fe). In the last hundreds of Myrs, however, the rate of SNe II is zero, but the SNe Ia rate is not negligible, so the gas that is recently 
lost from the galaxy is enriched mainly by SNe Ia products \citep[five times more Fe than O; ][]{Iwamoto99,Matteucci03}.

\section{Discussion}
\label{disc}

While we detected  an elongated structure in the centre of the NGC4325 group in the iron distribution,
the X-ray temperature map does not show any evidence of a relevant substructure.
This filament  was also confirmed by the metallicity profile extracted in identical regions
along and across this structure in the \textit{Chandra} and XMM-\textit{Newton} data.

Possible reasons for not observing the feature in the temperature map are the difficulty in observing 
an anti-correlation between temperature and metallicity because of projection effects \citep[as suggested in ][]{lovisari11}
and the fact that the gas thermalize faster than the mixing. In this way, metal inhomogeneities are observed 
for a longer time than temperature substructures.  We can also mention that the mean temperature of NGC4325 is about 1 keV, 
which is close to the temperature of the gas ejected from the galaxy, which makes the detection of any temperature 
feature almost impossible to be observed with current instruments.

This metal-rich structure in the core of the NGC~4325 group is spatially correlated with the central galaxy, and
because of its elongated shape, this structure could be due to any AGN activity. 
However, there is no evidence of nuclear activity in the X-ray images,  such as cavities, of the presence of a compact central source.
Also, the pseudo pressure and entropy maps do
not show any evidence of non-thermal heating processes of the IGM in the central region.
It is worth mentioning that we may have not detected the cavities because of the poor XMM-\textit{Newton} PSF and the
fact that this is not a deep observation of the NGC4325 group. 
Also, pseudo-entropy and pseudo-pressure maps depend on temperature, and as
mentioned before, it is hard to observe temperature inhomogeneities for this group.

Further support comes from 
the stellar population analysis of the optical spectrum of the central galaxy that showed no contribution of any recent AGN activity. 
Moreover, there is no detectable radio emission
associated with the central galaxy: no radio source was detected in the FIRST survey (down to 0.99 $\rm mJy beam^{-1}$), 
and none was found in the VLA all-sky survey (NVSS) within 1 arc min of the central X-ray emission. 
All this evidence indicate that there is no active AGN in the central galaxy.
However, it does not exclude 
any outburst of a 
previous AGN that is now too  
weak  to be detected. 

By means of chemical evolution models, 
we predicted the amount of gas and the mass of oxygen and of iron that this galaxy has lost through galactic winds.
Comparing these values to the observed oxygen and iron mass in the IGM, we can analyse 
the contribution of the central galaxy to the IGM enrichment and also the role of this mechanisms.
For this purpose, we computed the
iron and oxygen mass as the product of their abundance  by the gas mass enclosed within $r_{\Delta}$ and by
the solar photospheric abundance by mass \citep[0.0134;][]{aspl09}:
\begin{equation}
M_{\rm Z} = M_{\rm gas} (< r_{\Delta})  \times Z \times 0.0134.
\end{equation}

We find that the total iron and oxygen mass enclosed within $r_{\rm 2500}$ are  $M_{\rm Fe} \sim 3.5 \times 10^{9} M_{\odot}$ and
$M_{\rm O} \sim 2.2 \times 10^{9} M_{\odot}$. 
Comparing these values to the ones derived by a chemical evolution model presented in Table \ref{tab:evmod2},
 and assuming that all the iron and oxygen synthesized 
in the central galaxy of the group NGC4325 during its lifetime were released into the IGM (these mass values are then upper limits), 
the winds from the central galaxy alone play a minor role in the metal enrichment of the IGM. It accounts for  no more than 3.0$\%$ 
of the Fe mass within $r_{\rm 2500}$ and about 17-21$\%$ of the oxygen mass within this radius.
If we consider the flat radial distribution of the oxygen shown in Fig.\ref{fig:FeOprof} with a projected metal-abundance of $\sim$0.35 solar, 
it indicates that the oxygen becomes well mixed through the IGM, supporting the scenario in which the 
enrichment by oxygen occurs in the earlier stages of the group formation.
From chemical evolution models, we showed that the wind in the central galaxy
starts at a galactic age between 0.22 and 0.40 Gyr,
injecting much more O than Fe due to the short timescale for the production of O by 
massive stars and posterior release into the interstellar medium (ISM) by SNe II.

Simulations \citep[e.g.][]{schindler05} also suggest that the O
 from SNe II may have been ejected in protogalactic winds \citep{LD75}.
During the life time of the group, the winds get  driven more and more by SNe Ia. Thus, 
as the evolution proceeds, the amount of Fe in the ISM and, consequently in the wind, gets higher owing to the 
increase in the number of  SNe Ia. For the central galaxy, we showed that in the last
hundreds of Myrs the rate of SNe II is zero, but SNe Ia rate is not negligible, so the gas that has recently 
been lost from the galaxy is enriched mainly by SNe Ia products like Fe. This contributes to the fact of observing an iron 
distribution that is centrally peaked (as shown in Fig.\ref{fig:FeOprof}).
The iron mass released by the central galaxy is about 20\% of the iron mass inside a region of 
$ r \sim$ 77 arcsec, which encompasses the filamentary structure.

As discussed above, the winds from the central galaxy alone play a minor role in the IGM enrichment inside $r_{2500}$.
Thus, to explain the metal-rich structure in the 
core of the groups NGC4325, the most probable scenario is 
past  AGN activity, and since iron is not expected to diffuse over long 
distances during the life time of a group, it still appears as inhomogeneities in the
metallicity map. The AGN activity has to have happened when the SNe Ia rate was already higher than SNe II to uplift more
Fe to the IGM. Looking to Figs.\ref{fig:snia} and \ref{fig:snii}, we see that at a galactic age of 13.5 Gyrs, the SNe II rate 
goes to zero (in BC03 model), or decreases substantially (in the other models) and the
SNe Ia products dominate the winds. Thus, the AGN have to be younger than 5 $\times 10^{8}$ yrs and older than
$10^{7}-10^{8}$ yrs \citep[the inter-outburst period for the AGN has to be similar to the cooling time; e. g., ][]{gaspari11}.

Analysing \textit{Chandra} data, \citet{russell07} found two possible cavities in the residual image of this group.
Unlike most cavities discovered previously, these features are not visible in the X-ray images, but in the residual maps. Thus,
these authors mention that these features  could also be caused by ellipticity in the data, rather than actual cavities, since no radio
emission is currently observed. If these cavities observed by \citet{russell07} provide evidence of past AGN activity, those authors
found the age of a few $10^{7}$ yrs., which is in line with our results.

Mergers  of member galaxies with the dominating central galaxy is another probable mechanism that could explain the metal-rich structure.
The hypothesis is that  recent mergers (e.g. $\sim$ 1Gyr age) may have enough power to disturb and displace the dark matter halo 
of the group entraining with the IGM.  
Thus, to complete our analysis, we analysed the overall galaxy distribution of the group NGC 4325 with respect to the central dominant galaxy.


\subsection{An overall view of the galaxies of the NGC4325 group} 

We examined the galaxy distribution of the NGC4325 group using spectroscopic samples of galaxies extracted from the DR10-SDSS. 
The sample comprises all objects with 
measured redshift classified as galaxies, located on a circular region of 2$^{\circ}$ radius centred on NGC4325 and out to the formal SDSS-limiting magnitude for spectroscopic observations, 
$r_{\rm lim} = 17.75^{m}$. This sample contains 1.398 galaxies and is 68\% complete when compared to a similarly extracted DR10 sample containing all photometric objects classified as galaxies, 
regardless of spectroscopy and out to the same magnitude limit (see discussion below).
 
\begin{figure}
\centering
\includegraphics[width=6.6cm,angle=-90]{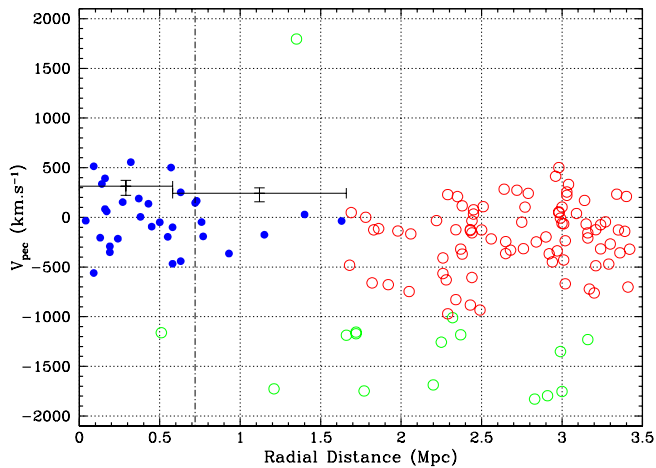}
    \caption{Phase space distribution of the galaxies of NGC4325. Blue solid dots show 
   the positions of the 32 galaxies found to be members of the system from the shifting gap technique. 
   The crosses give the velocity dispersion within each of the two rings in which the SGT divided the 
   sample for radial analysis. The horizontal error bars are equal to the width of the rings. 
   The open circles show the phase space positions of the the remnant 99 galaxies of the original 
   $|v_{pec}|\le 2000 \rm~km~s^{-1}$ sample, red  for galaxies with $|v_{pec}| < 1000\rm~km~s^{-1}$, and green for 
the $|v_{pec}| \ge 1000\rm~km~s^{-1} $ galaxies. The vertical dashed line marks the position of the virial 
radius of the cluster, $R_{200}$ (see text). }
 \label{fig:Vrad}
\end{figure}

\begin{figure}
\centering
\includegraphics[width=7.6cm]{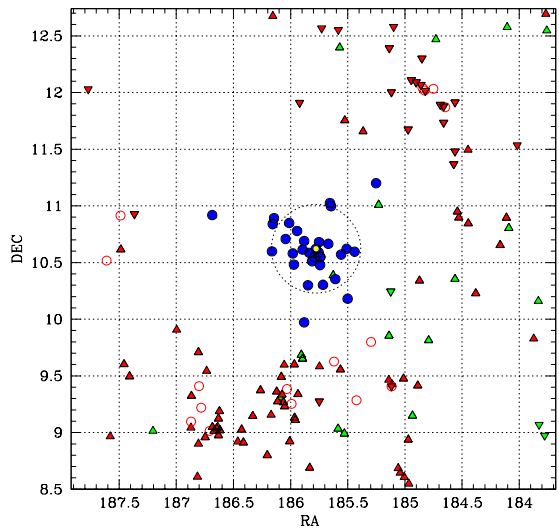}
   \caption{Projected sky positions of galaxies having peculiar velocities $V_{pec}\le 2000 \rm~km~s^{-1}$. 
    Colours follow Figure \ref{fig:Vrad}. The dominant galaxy NGC4325 is indicated by an extra 
    yellow circle at its position. Galaxies not belonging to the group (red and green) are 
    differentiated according to whether their speeds are lower (open circles) or greater 
    (triangles) than $100\rm~km~s^{-1}$ and  according whether they have negative peculiar velocities 
    (normal triangles) or positive (inverted triangles). 
    The virial radius $R_{200} = $ 0.69Mpc is shown by the dotted circle centred at NGC4325.}
\label{fig:positions}
\end{figure}

The detection of the group and the selection of galaxies kinematically belonging to it was made by applying the shifting gap technique \citep[SGT, as implemented by][]{Lopes09,Ribeiro13}.
This technique allows for the expected decrease in the velocity dispersion of the system with the radial distance and has proven to be efficient at removing outliers 
from the projected galaxy distribution. The final list of member galaxies, that is, those kinematically linked to the group, was found to have 32 members. 
Figure \ref{fig:Vrad}  shows the 
radial distribution of the peculiar velocities of all galaxies with $|v_{\rm pec}| \le 2000 \rm~km~s^{-1}$ on a $3.42$ Mpc circular region centred on the dominant elliptical galaxy 
NGC4325\footnote{This value corresponds to 1.94$^{\circ}$ for the cosmology adopted here.}. 
Peculiar velocities are calculated as $v_{\rm pec}=\rm c(z - z_{\rm group})/(1+z_{\rm group})$, where $z$ is the redshift of the galaxy and $z_{\rm group}$ 
the mean redshift of the system. As can be seen from this figure, the group members tend to concentrate on a region of $\sim 1$Mpc radius, 
with a sparse extension out to $1.8$ Mpc.  We notice the presence of 
a secondary structure in phase space at nearly the same redshift as the central group. These galaxies are not recognized as group members by the SGT essentially 
because of $\sim 0.5$Mpc radial gap existing between the two main structures. 
In Figure \ref{fig:positions}, it can be seen that these galaxies are distributed in real space as a ring 
around the the group centre. 

\begin{figure}
\centering
\includegraphics[width=6.6cm,angle=-90]{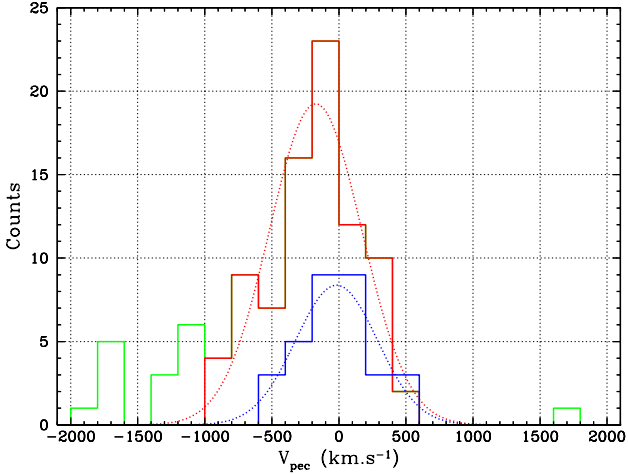}
    \caption{Histograms showing the velocity distributions of members (blue line) 
    and non-members galaxies (red and green lines, with the same colour scheme of Figure 
    \ref{fig:Vrad}). The dotted curves display the  corresponding Gaussian distributions.}
\label{fig:Vdistr}
\end{figure}

Figure \ref{fig:Vdistr} shows the velocity distribution of the galaxies shown in Fig. \ref{fig:Vrad}. We estimated the velocity dispersion using the scale robust bi-weighted 
estimator (Beer et al, 1990). We find $\sigma_{v}= 298^{-62}_{+78}~km~s^{-1}$. As seen, the velocity distribution of the group galaxies is well adjusted by a 
Gaussian with the velocity dispersion that we measured. This implies the characteristic radius for the virialised group \citep{Carlberg97} of $R_{\rm 200} = 720 {h_{70}}^{-1}$ kpc, 
($0^{\circ}.39$ angular radius) and 
comprises almost all the group with the exception of the sparse extension seen in Fig. \ref{fig:Vrad}.  

The low-velocity dispersion of the group is close to the internal velocity dispersion of the majority of normal galaxies. This is in fact a necessary 
condition for successful  mergers between galaxies (cf. Binney \& Tremaine, 1987), and it suggests a scenario where the group is evolving through successive mergers between its members. 

The bi-weighted mean velocity of the group is $V_{\rm bw} = 7641^{-94}_{+99} ~\rm km~s^{-1}$, to be compared with that of NGC4325, $V_{4325} = 7676 \rm ~km~s^{-1}$. This places the group 
at redshift $z_{\rm group} = 0.025488$.  From the kinematics of galaxies, we may estimate the dynamical mass of the system. We find $M_{vir} = 3.2^{+3.3}_{-1.6} \times10^{13}M_{\odot}$ 
for the usual Virial estimate \citep[e.g.][]{Girardi98}. 


Figure \ref{fig:positions} displays the map of projected positions of galaxies discussed in Fig. \ref{fig:Vrad}. The colours correspond to intervals of values of peculiar velocities, 
relative to mean recession velocity of the group. The group itself appears as the main concentration at the centre of the map. Remarkable in this figure are: $i$) the elongated 
structure of the NGC4325 group along the NW-SE direction, which although not exactly coincident with that of elliptical dominant NGC4325 galaxy (N-S), is not far from it (no obvious velocity gradient can be observed at the scale of the group); $ii$) the presence of galaxy concentrations other than the one at the centre and located almost on the borders of the field. 
These structures displays a noticeable gradient of the velocity gradient along the NE-SW direction, which is almost orthogonal  to the axis of the central group. 
It is very clear 
that these structures are not bound to the NGC4325 group, since they are too far by more than two times $R_{\rm 200}$ to have any influence on the central galaxies. 
However, it is tempting to 
hypothesise that they are all parts of a unique super structure, still to be studied, but seemingly centred at the NGC4325 group and still in its first collapse phase.   

Although we could not find any clear-cut evidence for recent mergers in the central parts of the NGC4325 (requiring, at least, detailed imaging and IFU 
observations of the central galaxy), we find suggestive signs that the system is still dynamically young and likely to evolve through mergers of its members.

\section{Conclusions}
\label{conc}
In summary, our analysis of the group NGC4325 led to the following conclusions:

\begin{itemize}

\item{The most striking feature in the core of NGC4325 is an elongated metal-rich filament structure that is spatially correlated with the central 
dominant galaxy of the group. }

\item{To confirm this structure, we extracted spectra in identical regions placed along and across the filament in the XMM-\textit{Newton}  and \textit{Chandra} data.
There is good agreement between XMM-\textit{Newton} and \textit{Chandra} values and a clear increase in the metallicity of the boxes
where the filament is located. This suggests that the filament is not a product of a poor spectral fit but it is a real structure of higher abundance.}

\item{The stellar population analysis of the optical spectrum and the analysis of the emission lines of the central galaxy showed 
no  recent AGN activity that could 
possibly lift metals up to the IGM, leading to the prominent feature revealed through the X-ray analysis. This result is not further supported by detectable
radio emission.}

\item{The analysis of emission line fluxes suggests shocks by SNe as the main ionization mechanism of the gas.}

\item{The stellar population analysis show that the central galaxy is dominated by an old stellar population (older than 2Gyr), but there seems to 
have been two more recent bursts.}

\item{Through chemical evolution models, we saw that a galactic wind is developed at early stages of galactic evolution, 
removing a fraction of the ISM of the galaxy and injecting the 
enriched gas into the IGM. In NGC 4325, the wind starts at a galactic age between 0.22 and 0.40 Gyr (depending on the model). 
As soon as the wind starts, the fraction of O that is injected into the IGM is higher than Fe thanks to the short timescale for the production of O by 
massive stars and subsequent release into the ISM by SNe II. As the evolution proceeds, however, the amount of Fe in the ISM and, consequently, in the winds 
gets higher owing to the increase in the number of supernova type Ia (SNe Ia). Since this type of SN originates in an intermediate mass star in a binary system, its timescale is longer.}

\item{In the last hundreds of Myrs, however, the rate of SNe II is zero, but the SNe Ia rate is not negligible, so the gas that is recently 
lost from the galaxy is mainly enriched  by SNe Ia products (like Fe)}


\item{The abundances observed in the IGM are the result of the production of the elements in stars and the transport of the elements released 
to their current location. Then, comparing the total iron and oxygen masses enclosed within $r_{2500}$ with the one release through SNe winds by 
the central galaxy, we conclude that the winds from the dominant galaxy alone play
a minor role in the IGM metal enrichment inside $r_{2500}$.}

\item{Analysing the metal profiles, we found that the iron abundance shows an increase towards the centre, while the oxygen abundance is radially more constant. 
Thus, Fe is still being added to the IGM specifically in the core by the SNIa, and O is a well-mixed product of the SNII in which the
enrichment occurs in the earlier stages of the group formation (as suggested by chemical evolution models).}

\item{To explain the elongated metal-rich filament structure in the core of this group, we suggested a past  AGN activity 
older than $\sim 10^7-10^8$ yrs and younger than $5 \times 10^{8}$ yrs. Since iron is not expected to diffuse over large 
distances during the life time of a group, it still appears as inhomogeneities in the
metallicity map.}

\item{Through the optical analysis of the overall galaxy distribution, we found no evidence of a recent merger in the centre of the group
that could explain the elongated structure, but we did find suggestive signs that the system is dynamically young. We noticed the presence of a secondary
structure in phase space at nearly the same redshift as the central group. This structure is distributed as a ring and about 0.5 Mpc of the group centre.
We hypothesised that they are all part of a common super structure still in its first collapse phase.}

\end{itemize}

\begin{acknowledgements}
We thank the anonymous referee for relevant suggestions that improved the quality of this manuscript.
TFL acknowledge financial support from FAPESP (grants: 2012/00578-0 and 2012/13251-9).
LL acknowledges support by the DFG through grant LO2009/1-1 and by the Transregional Collaborative Research Centre TRR33 
''The Dark Universe" (project B18). LM thanks the CNPq for financial support through grant 305291/2012-2, and 
GS acknowledges support by the German Research 
Association (DFG) through grant RE 1462/6.

This research is based on observations obtained with XMM-\textit{Newton}, an ESA science mission 
with instruments and contributions directly funded by 
ESA Member States and NASA. This research also made use of data obtained from the \textit{Chandra} Data Archive and of software
provided by the \textit{Chandra} X-ray Center (CXC) in the application packages \texttt{Ciao}.

Funding for SDSS-III has been provided by the Alfred P. Sloan Foundation, the Participating Institutions, the National Science Foundation, and the U.S. Department of Energy Office of Science. The SDSS-III web site is http://www.sdss3.org/.
SDSS-III is managed by the Astrophysical Research Consortium for the Participating Institutions of the SDSS-III Collaboration including the University of Arizona, the Brazilian Participation Group, Brookhaven National Laboratory, Carnegie Mellon University, University of Florida, the French Participation Group, the German Participation Group, Harvard University, the Instituto de Astrofisica de Canarias, the Michigan State/Notre Dame/JINA Participation Group, Johns Hopkins University, Lawrence Berkeley National Laboratory, Max Planck Institute for Astrophysics, Max Planck Institute for Extraterrestrial Physics, New Mexico State University, New York University, Ohio State University, Pennsylvania State University, University of Portsmouth, Princeton University, the Spanish Participation Group, University of Tokyo, University of Utah, Vanderbilt University, University of Virginia, University of Washington, and Yale University.
\end{acknowledgements}

\bibliographystyle{aa} 
\include{adsjournalnames} 
\bibliography{refs} 

\appendix

\section{Error maps}

In Fig.A.1. we present the errors associated to each bin in the temperature and metallicity maps displayed in Fig.3.
The error on these parameters was obtained directly from the spectral fits. 
For all the analysed spectra the metallicity and temperature errors are smaller than 3\% and 15\%, respectively. 
Thanks to this good accuracy and together with the profiles showed in Fig. 5 we can conclude that the high-metallicity structure is indeed real.

\begin{figure}[ht]
\centering
\includegraphics[scale=0.4]{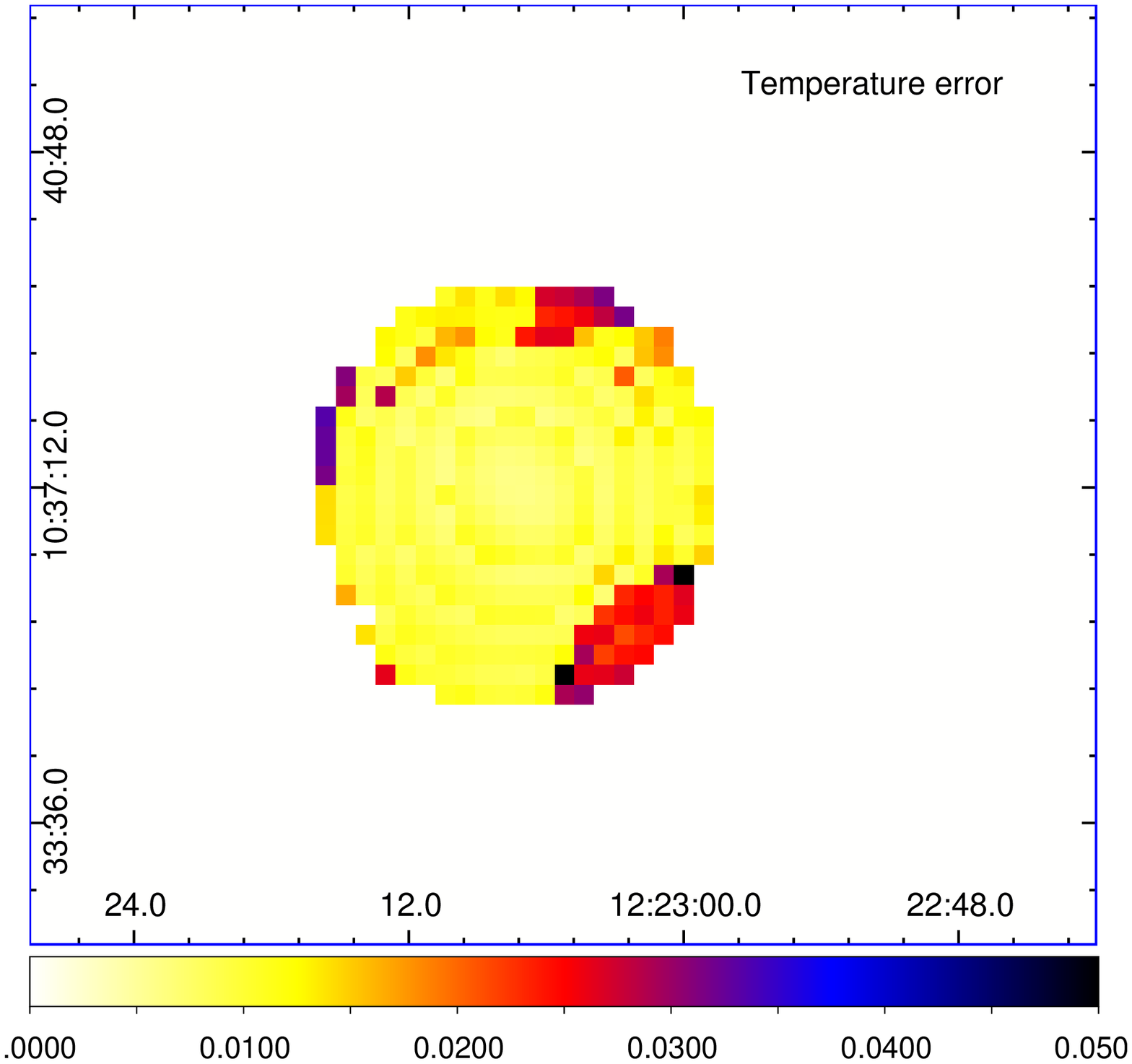}
\includegraphics[scale=0.4]{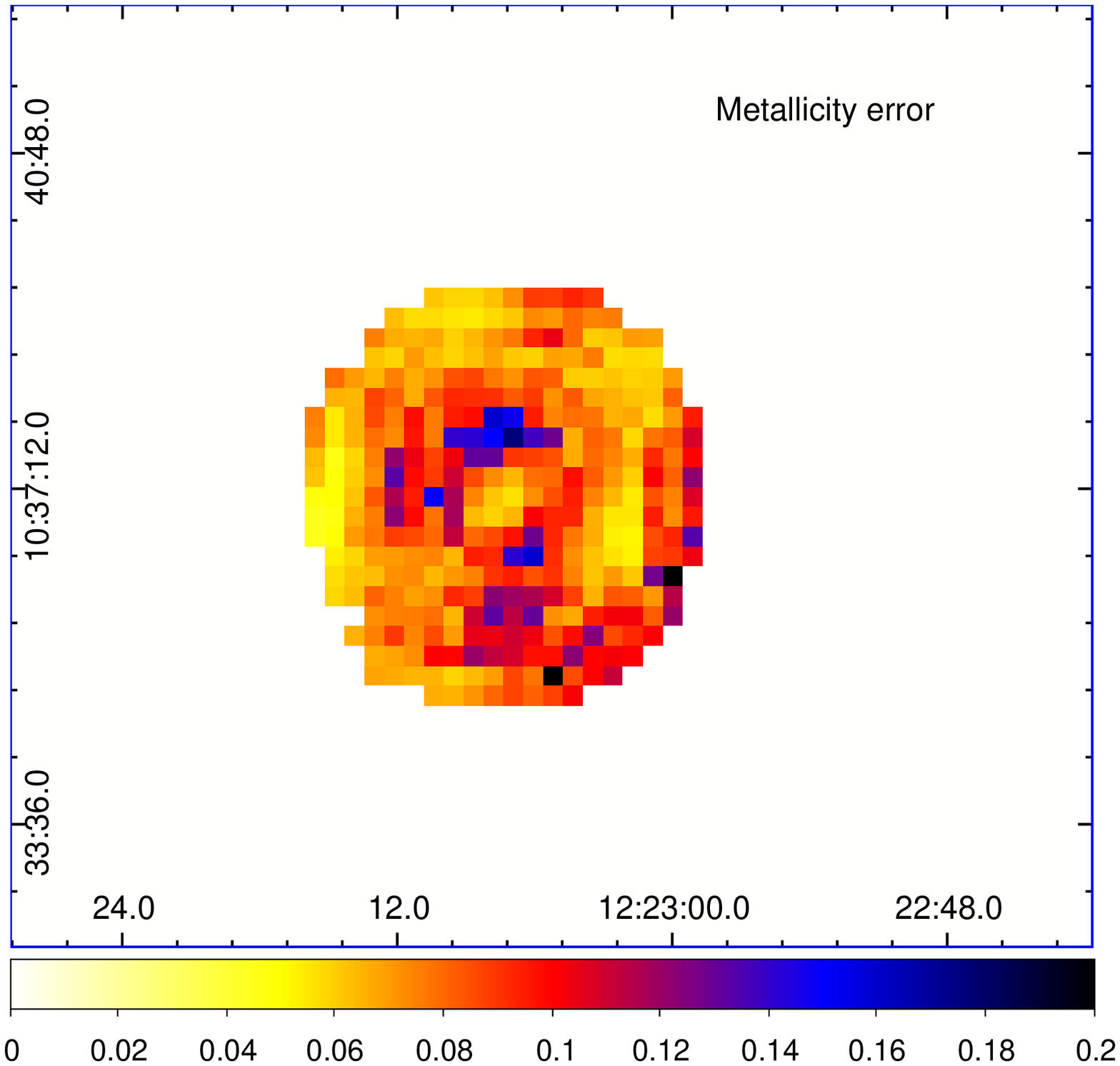}
\caption{Upper panel: temperature error map. Lower panel: metallicity error map.}
\label{fig:err_maps}
\end{figure}

\end{document}